\documentclass[]{article}

\usepackage{arxiv}
\usepackage{caption}
\usepackage[utf8]{inputenc} 
\usepackage[T1]{fontenc}    
\usepackage{hyperref}       
\usepackage{url}            
\usepackage{booktabs}       
\usepackage{amsfonts}       
\usepackage{nicefrac}       
\usepackage{microtype}      
\usepackage{lipsum}
\usepackage[super,sort&compress,comma]{natbib}

\usepackage{graphicx}   

\usepackage{url}        

\usepackage{amsmath}    

 \usepackage{datetime}
 
\usepackage{tikz}
\usepackage[english]{babel}
\usepackage{graphicx}
\usepackage{amsthm,color}
\title{Stiffening and Inelastic Fluidization in Vimentin Intermediate Filament Networks}

\author{
  Anders Aufderhorst-Roberts\thanks{Present address:\textit{~School of Physics and Astronomy, University of Leeds, Leeds LS2 9JT, United Kingdom.}}\\
  Living Matter Department\\
  AMOLF\\
  1098 XG Amsterdam\\
  The Netherlands\\
  \texttt{a.aufderhorst-roberts@leeds.ac.uk} \\
   \And
 Gijsje H. Koenderink \\
  Living Matter Department\\
  AMOLF\\
  1098 XG Amsterdam\\
  The Netherlands\\
  \texttt{g.koenderink@amolf.nl} \\
}

\begin{document}
\maketitle

\begin{abstract}
Intermediate filaments are cytoskeletal proteins that are key regulators of cell mechanics, a role which is intrinsically tied to their hierarchical structure and their unique ability to accommodate large axial strains. However, how the single-filament response to applied strains translates to networks remains unclear, particularly with regards to the crosslinking role played by the filaments' disordered ``tail'' domains. Here we test the role of these noncovalent crosslinks in the nonlinear rheology of reconstituted networks of the intermediate filament protein vimentin, probing their stress- and rate- dependent mechanics. Similarly to previous studies we observe elastic stress-stiffening but unlike previous work we identify a characteristic yield stress $\sigma^*$, above which the networks exhibit rate-dependent softening of the network, referred to as \textit{inelastic fluidization}. By investigating networks formed from tail-truncated vimentin, in which noncovalent crosslinking is suppressed, and glutaraldehyde-treated vimentin, in which crosslinking is made permanent, we show that rate-dependent inelastic fluidization is a direct consequence of vimentin's transient crosslinking. Surprisingly, although the tail-tail crosslinks are individually weak, the effective timescale for stress relaxation of the network exceeds 1000s at $\sigma^*$. Vimentin networks can therefore maintain their integrity over a large range of strains (up to $\sim$1000$\%$) and loading rates (10$^{-3}$ to 10$^{3}s^{-1}$). Our results provide insight into how the hierarchical structure of vimentin networks contributes to the cell's ability to be deformable yet strong.  
\end{abstract}

\footnotetext{\dag~Electronic Supplementary Information (ESI) available: Onset stress and linear modulus at all loading rates (S1), Representative stress vs. strain curve indicating rupture criteria (S2), loading rate dependent contributions to differential storage modulus (S3), affine entropic simulation data of vimentin networks (S4), frequency sweeps of vimentin networks with different crosslink modifications (S5), linear and nonlinear rheology of tailless vimentin (S6).}

\twocolumn
\section{Introduction}
The cells in our body have the remarkable ability to maintain but also adjust their structure in dynamic and stressful environments.\cite{burla2018biopolymers} This remarkable versatility is facilitated by the cytoskeleton, an intracellular filamentous network that spans the cell's interior. Cells are able to stiffen in response to mechanical cues\cite{wang1994control,pourati1998cytoskeletal,fernandez2006master,kollmannsberger2011nonlinear}
but have also been shown to soften\cite{trepat2007universal,krishnan2009reinforcement,chen2010fluidization} depending on the magnitude of applied stress,\cite{gardel2006stress} the loading rate \cite{kasza2010actin} and the contributions of different cytoskeletal components \cite{esue2006direct} and the interactions between them.\cite{huber2015cytoskeletal} 

\begin{figure*}[ht]\centering 
\includegraphics[width=\linewidth,trim={0 -1pt 0 0},clip]{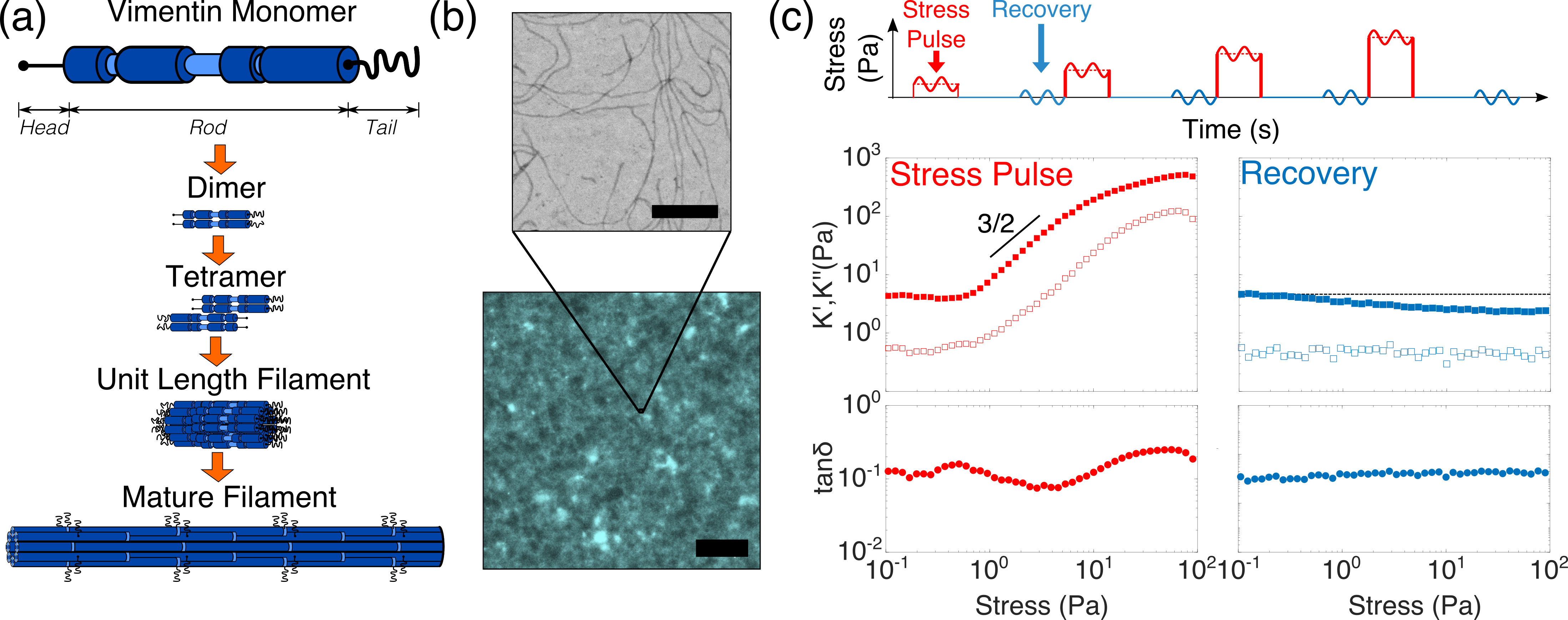}
\caption{(a) Vimentin monomers comprise a central rod domain of three $\alpha$-helical coils with disordered ``head'' and ``tail'' regions at either end.  Monomers assemble from parallel coiled-coil dimers into full-length filaments through a series of intermediate steps. (b) In the presence of monovalent and divalent cations, vimentin assembles into smooth filaments (electron micrograph, top, scale bar: 0.5$\mu$m) that form dense networks (confocal fluorescence micrograph, bottom, scale bar: 20$\mu$m). (c) Here the ``stress pulse'' rheology protocol short stress pulses are interspersed with longer recovery periods. Stress pulses show extensive stiffening, a consequence of the uniquely high rupture strains of intermediate filaments.  Networks largely recover upon the removal of stress. }
\label{fig:stresspulse}
\end{figure*}

An integral subsystem of the cytoskeleton are the intermediate filaments, which form cell-spanning networks that are essential to mechanotransduction and provide resistance to external loads.\cite{guo2013role} Intermediate filament networks are very soft at small deformations but are able to stiffen substantially under stress\cite{herrmann2007intermediate} and can withstand significantly higher axial strains than the other two cytoskeletal filaments, actin filaments and microtubules. \cite{janmey1991viscoelastic} These unusual mechanical properties are thought to be related to the extensible $\alpha$-helical domains present in all intermediate filament monomers and the precise assembly sequence from monomers to filaments under physiological conditions\cite{strelkov2003molecular} (figure \ref{fig:stresspulse}(a)). Their primary role as mechanoprotective elements is evidenced by the tendency of  many cell types to increase their expression levels of intermediate filaments in response to stress\cite{flitney2009insights} and the numerous intermediate filament genetic mutations associated with diseases relating to cell fragility. \cite{omary2004intermediate} They are expressed in a cell- and organism- specific manner, suggesting their physiological function may relate directly to the local mechanical environment in different tissues. 

Intermediate filament networks exhibit dynamic (dis)assembly and remodeling in the cell,\cite{robert2016intermediate} but on a much larger timescale (minutes) compared to the turnover dynamics of  actin filaments and microtubules (seconds), which enables them to act as long-lived structures that help maintain cell integrity\cite{harris2012characterizing,latorre2018active} and cytoskeletal polarity\cite{jiu2017vimentin} even as cells perform dynamic processes such as migration.\cite{de2018intermediate,gan2016vimentin} 
How the interplay between dynamic remodelling of intermediate filaments and their nonlinear mechanics depends on the mechanical load and the loading rate remains an unresolved question. In cells, interactions between intermediate filaments are mediated by a variety of accessory proteins that crosslink the filaments to each other,\cite{wiche2011plectin} to actin and microtubules,\cite{coulombe2000ins} and to cell-matrix and cell-cell adhesions.\cite{quinlan2017rim}  

Rheological studies on reconstituted networks of intermediate filaments have furthermore shown that they are inherently associative due to ionic interactions, \cite{pawelzyk2014attractive} particularly between their highly charged carboxy-terminal tail domains which act as crosslinkers. \cite{lin2010origins} As a result of this crosslinking, intermediate filament networks exhibit nonlinear stress stiffening over several decades of applied stress.\cite{janmey1991viscoelastic} \cite{pawelzyk2014attractive,schopferer2009desmin} The non-covalent nature of these interfilament interactions \cite{janmey2014polyelectrolyte} presents a potential mechanism for dissipation, through a process known as \textit{inelastic fluidization}.\cite{gralka2015inelastic} In contrast to entropic stiffening, which arises from the initial nonlinear stretch response of individual filaments,\cite{broedersz2014modeling} inelastic fluidization emerges from subsequent remodelling events such as  unbinding and rebinding of the crosslinks between filaments leading to network softening. This behaviour was first observed in reconstituted actin networks,\cite{wolff2012resolving,lieleg2008transient} in which  transiently bound accessory proteins act as crosslinkers between filaments.\cite{lieleg2009cytoskeletal} Observing inelastic fluidization during mechanical loading is  technically challenging because time-dependent and stress dependent  mechanics are inherently convoluted in nonlinear rheology experiments.\cite{broedersz2010measurement} Moreover, it is challenging to discriminate inelastic effects at the network level from inelastic effects at the filament level.  For intermediate filaments, these include rate-dependent extensibility\cite{block2017nonlinear} and dissipation\cite{block2018viscoelastic} as measured by single-filament stretching experiments.

Here we investigate the consequences of noncovalent interactions for the nonlinear rheology of vimentin networks, using two complementary rheology protocols that allow us to deconvolve the time and stress-dependent mechanics. We are thus able to extract quantitative measures of the time scales and characteristic stress where inelastic fluidization occurs. We directly probe the contribution from noncovalent interactions between filaments by suppressing these interactions through genetic modification by removing the C-terminal tails and through chemical crosslinking. This comprehensive approach allows us to construct a phase space describing how the nonlinear response depends on loading rate and amplitude. 
\begin{figure*}[ht]\centering 
\includegraphics[width=\linewidth]{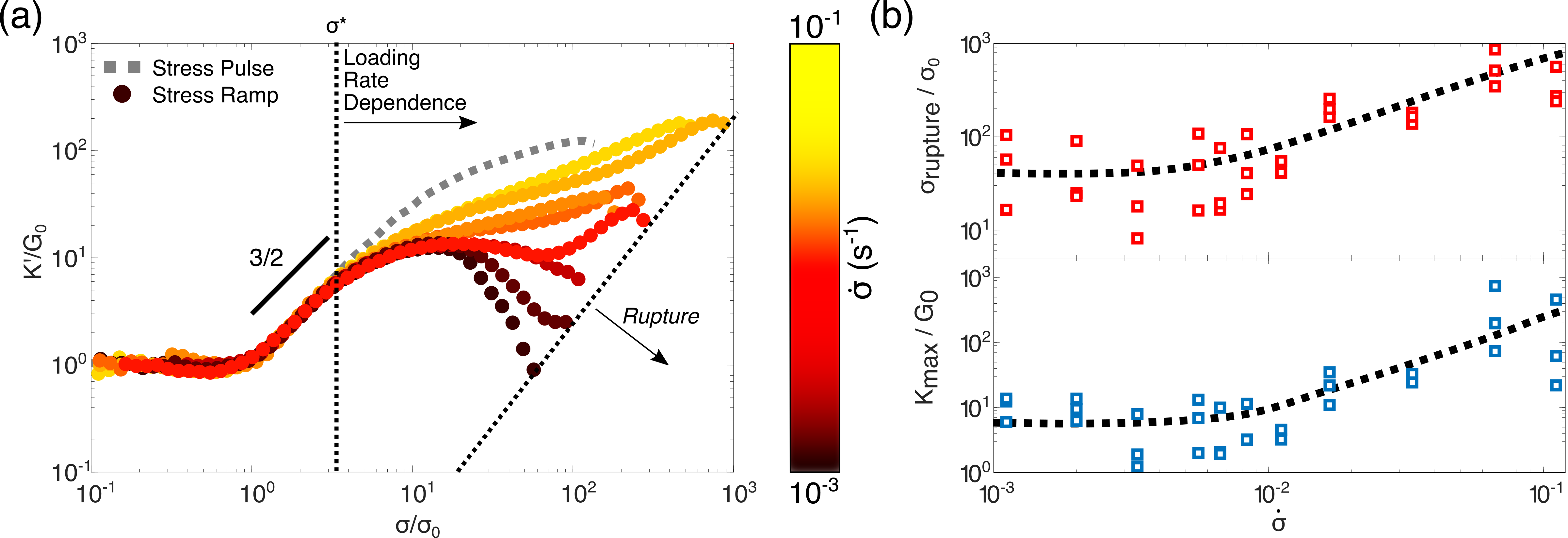}
\caption{Rate dependent mechanics of vimentin intermediate filament networks. (a) Logarithmic stress ramps (circles) at variable loading rates, $\dot{\sigma}$, to give the differential storage modulus K' with respect to stress $\sigma$ normalised by linear modulus $G_0$ and onset stress $\sigma_0$. The response shows close agreement with the stress pulse protocol (grey dashed line) below a threshold stress ${\sigma}^*\approx2.5\times\sigma_0$ (vertical dashed line) but is highly dependent on loading rate above $\sigma^*$.  (b) Faster loading rates correspond to increased rupture stress (top) and increased peak storage modulus (bottom). Lines are guides to the eye. }
\label{fig:stressramps}
\end{figure*}
\section*{Methods}
\addcontentsline{toc}{section}{Methods}
\subsection*{Protein Preparation}
Human vimentin was expressed in Escherichia coli (strain TG1) and purified from inclusion bodies as previously described.\cite{herrmann1992identification}  A tailless mutant of human vimentin, with a truncation of the final 55 residues of the carboxy-terminal,\cite{herrmann1996structure} was purified separately. Plasmid DNA for wild type and tailless vimentin were gifted by H. Herrmann, German Cancer Research Centre. Purified vimentin protein was stored at -80$^\circ$C in a storage buffer comprising 8 M urea, 5 mM Tris-HCl, pH 7.5, 1 mM dithiothreitol (DTT), 1 mM ethylenediaminetetraacetic acid (EDTA), 0.1 mM ethylene glycol bis($\beta$-aminoethyl ether)N,N'-tetraacetic acid (EGTA). The protein was dialyzed from 8M urea in steps of decreasing urea concentrations, (6M, 4M, 2M, 1M) by dilution with 5 mM Tris-HCl, pH 8.4, 1 mM
EDTA, 0.1 mM EGTA, and 1 mM DTT, at room temperature with dialysis tubing with a molecular weight cut off of 12,000-14,000 as previously described\cite{herrmann2004isolation}. Dialysis was continued overnight at 4$^\circ$C against a final buffer of 5 mM piperazine-N,N'-bis(2-ethanesulfonic acid) (PIPES), pH 7.0, 1 mM EGTA and 1 mM DTT. The final protein concentration was determined through measurement of the UV absorbance at 280nm using extinction coefficients of 22450M$^{-1}$cm$^{-1}$ for wild type vimentin and  24870$^{-1}$cm$^{-1}$ for tailless vimentin. The scattering contribution to the UV spectra was measured by fitting a Lambert-Beer law within the range 320-340nm and subtracting the result from the measured spectra.\cite{birdsall1983correction} Dialysed protein was aliquotted, flash-frozen in liquid nitrogen and stored at -80$^\circ$C. To initiate filament assembly into networks we added a 10$\times$ concentrated polymerization buffer to thawed vimentin at 37$^\circ$C to obtain final concentrations of 25mM PIPES (pH 7.0), 100mM KCl, 4mM MgCl$_2$ and 1mM EGTA.  

\subsection*{Shear Rheology}
All rheology measurements were performed with a stress-controlled rheometer (Kinexus Malvern Pro) equipped with a steel cone and plate (20mm diameter, 1$^\circ$ cone angle). Immediately after mixing with polymerization buffer the sample was transferred to the rheometer. Samples were left to polymerize for 60mins at 37$^\circ$C between the rheometer plates by which time the storage modulus G' was observed to reach a steady state value. Mineral oil (Sigma Aldrich) was applied to the air-sample interface to prevent drying.  Vimentin was polymerised at a concentration of 1mg/ml, unless otherwise stated.  Assuming a mass per length of 6.65$\times10^{-11}$g/m,\cite{herrmann1999characterization} the mesh size of the network $\xi$ is estimated by the relation $\xi = \frac{1}{\sqrt{\rho}}$, where $\rho$ is
the filament density expressed in terms of the total filament length per volume.  A concentration of 1mg/ml of wild type vimentin with molecular weight 53.6kDa corresponds to $\xi$ = 206nm. 

We term the two main nonlinear rheology protocols in this work the ``stress pulse'' and ``stress ramp'' protocols.  In the stress pulse protocol, a steady prestress was applied for 20s, superimposed with a small amplitude oscillatory stress $\delta\sigma(t) = \delta\sigma e^{i \omega t}$.  The resulting oscillatory strain response $\delta\gamma (t)  = \delta\gamma e^{i\omega t}$ was measured  at a frequency of 0.5Hz. The oscillatory stress was fixed at $10\%$ of the steady prestress and the first 10s of oscillations were discarded to eliminate the possible influence of instrument inertial contributions.  The stress was set to 0Pa for a period of 60s between successive measurements to probe network recovery following perturbation. The complex differential modulus  in this approach, $K(\omega,\sigma)=\delta\sigma/\delta\gamma$, is  an instantaneous measurement\cite{broedersz2010measurement} and independent of inelastic fluidization over longer timescales. From the complex modulus, we derive the differential storage modulus, (K') the differential loss  modulus (K'') and the loss tangent (K''/K'). By contrast, in the stress ramp protocol,\cite{lieleg2007cross,semmrich2008nonlinear} no  oscillations were superimposed, and the sample strain was measured while the steady stress was  increased at a fixed logarithmic rate $\dot{\sigma}$, defined in units of decades of applied stress per second. The resulting differential modulus K' was calculated by applying a numerical derivative to the stress-strain curve. In the stress ramp protocol, K' is dependent both on the elastic response of the constituent filaments and any inelastic fluidization, as demonstrated in earlier studies of entangled actin networks.\cite{broedersz2010measurement,semmrich2008nonlinear} 

The time-dependence of network deformation over longer time scales was probed by a creep test, in which a constant shear stress was applied for 600 seconds and the resulting strain was measured.  The creep rate was determined by performing a linear fit over the final 60s of the time dependent strain. The timescale for stress relaxation (where the elastic and viscous moduli cross-over) were probed by frequency-dependent oscillatory measurements over a frequency range of 10$^{-3}-1$Hz, with a strain amplitude of 0.5$\%$. When probing the frequency response in the nonlinear regime, oscillations were superimposed on a steady stress through the same principles as outlined in the stress pulse protocol above.  Repeated frequency sweeps were applied to each sample to verify that the samples were unaltered during the course of the measurement. 

\subsection*{Fluorescence Imaging}
\addcontentsline{toc}{subsection}{Fluorescence Imaging}
Wild type vimentin was labelled through conjugation of its single cystein residue with Alexa Fluor 488 maleimide (Invitrogen), as previously described.\cite{winheim2011deconstructing} Labelled and unlabelled vimentin were mixed in storage buffer before dialysis, with a fraction of labelled vimentin of approximately 5$\%$. Imaging was carried out using a Nikon A1 confocal microscope with a perfect focus system, a 100x/NA1.4 oil immersion objective,
and a 100-mW 488 nm argon ion laser.

\subsection*{Scanning Transmission Electron Microscopy}
\addcontentsline{toc}{subsection}{Scanning Transmission Electron Microscopy}
Imaging was performed as previously described\cite{lopez2018effect} on a Verios 460 electron microscope (FEI) operating in bright field mode at an acceleration voltage of 20 kV. Vimentin filaments were assembled at a concentration of 0.01mg/ml for 1 hour and deposited on carbon-coated copper grids (Ted Pella). After 1 minute of deposition, the grids were rinsed with assembly buffer and adsorbed filaments were fixated by incubating with 0.1$\%$(w/v) glutaraldehyde dissolved in Milli-Q water (Merck) for 5 minutes. Before imaging, grids were rinsed 5 times with Milli-Q water and air-dried.

\subsection*{Network Modelling}
In order to calculate the expected rheological response in the limit of affine (uniform) network deformation, we computed the stress-strain response of isotropic ensembles of non-interacting rigid rods. Isotropic ensembles of filaments were generated by a custom-written Matlab code. A total of 10$^4$ filaments of length $l_c$ were randomly deposited within a three-dimensional cube. The value of $l_c$ was inferred from measurements of the linear storage modulus $G_0$ using the relation:\cite{mackintosh1995elasticity} $l_c = \left(\frac{6\rho k_B T l_p^2}{G_0}\right)^{1/3}$ where $k_B$ is the Boltzmann constant, T is the experimental temperature of 310K and $l_p$ is the persistence length of vimentin (2.1$\mu$m\cite{noding2012intermediate}). The length density of filaments in the simulation was selected to match the experimental length per volume $\rho$. Each filament was assigned a random orientation and subjected to a homogeneous shear strain. The resulting filament tension was computed through an analytical expression for the force-extension relation of an inextensible semiflexible polymer.\cite{broedersz2014modeling}

\section*{Results}
We first compare the nonlinear viscoelastic response of vimentin networks through two rheology protocols.  Figure \ref{fig:stresspulse}(c) shows the results from the ``stress pulse'' protocol which probes the elasticity of the network in response to an instantaneous stress.\cite{gardel2004elastic} As the stress is increased above a threshold stress $\sigma_0$, both the differential storage modulus K' and loss modulus K'' are observed to increase, indicating network stiffening. K' increases with a  power law of exponent 3/2, consistent with models of entropically-driven stiffening of semiflexible polymers and indicating that the network response inherits the force extension behaviour of its constituent filaments.\cite{mackintosh1995elasticity,broedersz2014modeling} At higher stress, the stiffening exponent begins to level-off, indicating the enthalpic stretching of the individual filaments. \cite{lin2010divalent, lin2010origins}

The dissipation of stress in the vimentin network is quantified by the loss tangent, tan$\delta =K''/K'$. The instantaneous nature of the stress pulse protocol means that dissipation has its origins in the viscous drag on the fluctuating filament segments.\cite{gittes1998dynamic} A slight increase in tan$\delta$ is observed immediately prior to the onset of stiffening, which we speculate may be due to the unbinding of weaker binding sites that have previously been reported to exist on vimentin's rod domain\cite{pawelzyk2014attractive}. In the stress regime where the network stiffens, tan$\delta$ decreases due to the damping of filament fluctuations.\cite{gittes1998dynamic} Only at high stress does  tan$\delta$ begin to increase, indicating progressive crosslink unbinding and incipient network fracture.  This leads to eventual network rupture, defined as tan$\delta >1 $, consistently measured as  $\sigma = 100Pa$ over 3 repeat experiments. We examine the plasticity of the network across these different regimes of applied stress by probing the mechanics following 60s of strain relaxation. A remarkably high degree of recovery of the original modulus of the virgin network is observed as,  even immediately prior to the rupture of the network, G' decreases no more than 2-fold in comparison to the original linear modulus. We note that this behaviour contrasts strongly with actin networks, which retain a strong ``memory'' of prior cycles of stress\cite{schmoller2010cyclic} and strain.\cite{majumdar2018mechanical}

So far, we have examined only the instantaneous elasticity of the vimentin networks.  To examine time-dependent mechanics we probe the same networks with the stress ramp protocol. Here, the measured mechanical response is the result of a convolution of two separate phenomena, the entropic stiffening previously shown in figure \ref{fig:stresspulse} and the inelastic fluidization of the vimentin network over time. Direct agreement between the protocols would therefore indicate the complete absence of inelastic fluidization.\cite{lieleg2007cross,broedersz2010measurement,semmrich2008nonlinear} Figure \ref{fig:stressramps}(a) shows the nonlinear response from the stress ramp protocol over a range of loading rates of 10$^{-1}$-10$^{-3}$s$^{-1}$, where the rate denotes decades of applied stress per second. At low applied stress, the initial stiffening as well as the linear modulus and onset stress (see figure S1) are all in close agreement with the entropic stiffening measured by the stress pulse protocol. By contrast, above a threshold stress $\sigma^*$ the nonlinear response depends strongly on the loading rate: with large deviations from the stress pulse protocol observed at the slowest loading rates, and more subtle deviations seen at the fastest loading rates. As a consequence, both the degree of stiffening (figure \ref{fig:stressramps} (b), bottom) and the rupture stress,(figure \ref{fig:stressramps} (b), top)
for convenience defined as the stress corresponding to a 5 orders of magnitude increase in strain (figure S2), increase substantially with loading rate. 

The loading-rate dependent nonlinear response indicates a softening of the network which counteracts the entropic stiffening that is barely perceptible at the fastest loading rates ($\dot{\sigma} > 0.1s^{-1}$) but clearly present at the slowest loading rates ($\dot{\sigma} < 0.1s^{-1}$). At intermediate loading rates ($\dot{\sigma} \approx 0.1s^{-1}$), 
we observe reentrant entropic stiffening, suggesting a sensitive interplay between entropic stiffening and softening.  Remarkably, the onset of inelastic fluidization at $\sigma>\sigma^*$ is independent of loading rate (figure S3). 

\begin{figure}[ht]\centering 
\includegraphics[width=\linewidth]{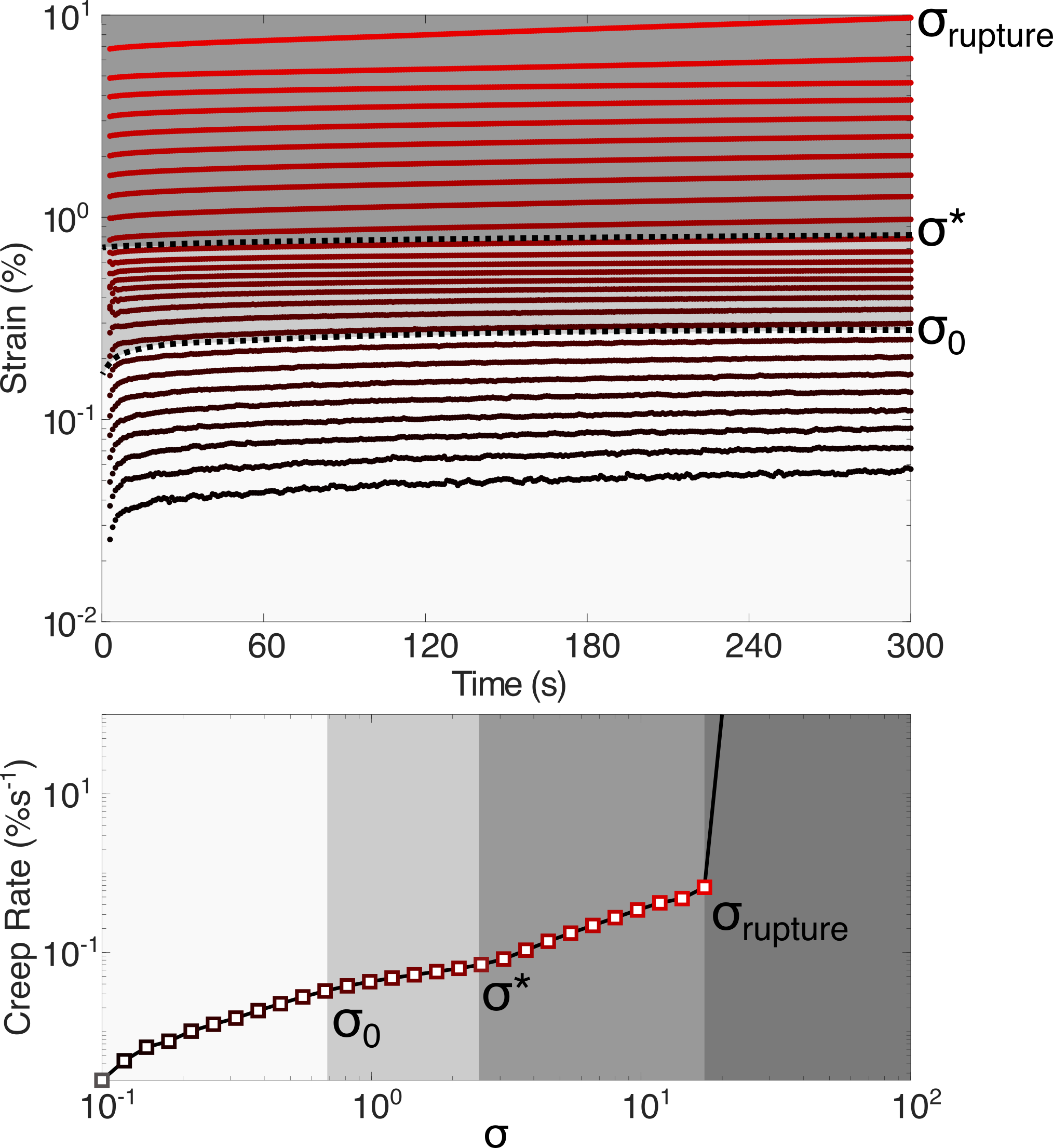}
\caption{Independent verification of vimentin network inelastic fluidization through the dependence of creep (a) and creep rate (b) on the level of applied stress. Consistent with stress ramp data (figure \ref{fig:stressramps}) the creep and creep rate both increase strongly above a threshold stress $\sigma^*$, indicating significant inelastic fluidization. Networks withstand strains up to 1000$\%$ at applied stresses of 20$\times\sigma_0$. }
\label{fig:creepcurves}
\end{figure}

To independently verify the stress dependence of inelastic fluidization, we carry out a series of creep experiments, measuring the time-dependent strain response of the network at different levels of stress. Figure \ref{fig:creepcurves} shows the time dependent strain at increasing applied stress (top) and the associated creep rate (bottom).   For $\sigma<\sigma_0$ the creep and creep rate both increase steadily with increasing stress, reaching a plateau at $\sigma\approx\sigma^*$.  At higher stress ($\sigma>\sigma^*$), a subsequent increase in creep and creep rate is observed, confirming the presence of a inelastic fluidization at $\sigma\approx\sigma^*$. Using the same definitions of the rupture stress $\sigma_{rupture}$ as for the stress ramp protocol we find $\sigma_{rupture}=20\times\sigma_0$ at applied strains of 1000$\%$, which corresponds approximately to the slowest loading rates in figure $\ref{fig:stressramps}$. 

The hierarchical structure of vimentin networks presents a number of possible mechanisms by which inelastic fluidization could occur at $\sigma>\sigma^{*}$. Firstly, it may be \textit{enthalpic}, arising from the stretching of individual filaments beyond their equilibrium contour length via the unfolding of coiled-coil rod domains\cite{qin2009hierarchical,block2017nonlinear}. To estimate the likelihood of this scenario, we calculate the expected forces exerted on individual filaments by modelling the network as an initially random ensemble of filaments that deforms affinely when a shear stress is applied, as shown in figure S4. The individual forces acting upon each filament at a given shear stress are determined by an analytical expression for the semi-flexible force-extension relation.\cite{broedersz2014modeling} The forces per filament are in the sub-piconewton range for  $\sigma<\sigma^*$.  At $\sigma\approx\sigma^*$ a proportion of filaments ($\approx15\%$) experience forces exceeding 1pN for $\sigma\approx\sigma^*$, but these forces are much lower than those reported to initiate axial stretch of the filaments (of order 100s pN)\cite{block2017nonlinear}). 

A more likely explanation is therefore that inelastic fluidization is caused by unbinding of crosslinks between the vimentin ``tail'' regions. The noncovalent nature of these crosslinks \cite{janmey2014polyelectrolyte} suggests a mechanism for unbinding under tension and would also be consistent with the observed loading rate dependence because the bond fraction depends not only on the maximum load applied over the course of the experiment but also on total loading time.

\begin{figure}[ht]\centering 
\includegraphics[width=\linewidth]{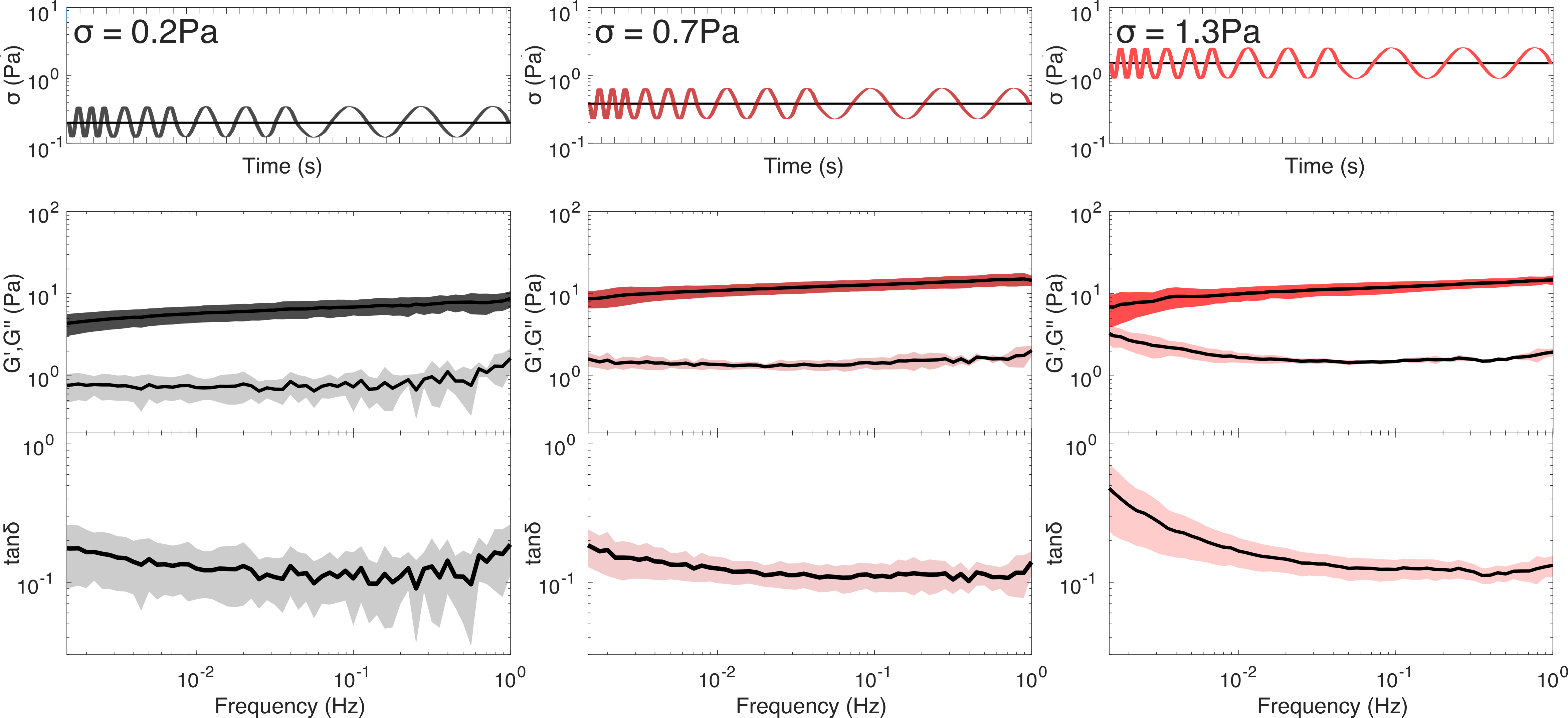}
\caption{Frequency sweeps of vimentin networks, averaged (N=3), with shaded regions representing standard deviation. Small amplitude oscillations are superimposed on to constant deformations simultaneously revealing stress and timescale dependence of moduli. At lower applied stress the moduli and loss tangent are largely invariant with frequency indicating that any characteristic timescales exceed the experimentally accessible range (1-1000s).  At higher applied stress, G' and G'' begin to converge at low frequencies indicating that the characteristic crosslink timescale decreases as applied stress is increased. }
\label{fig:freqsweeps}
\end{figure}

To measure the stress-dependent crosslink unbinding time, $\tau_{off}$  we perform oscillatory frequency sweeps  while simultaneously applying a constant stress, as shown in figure \ref{fig:freqsweeps}. We expect stress relaxation at frequencies below $1/\tau_{off}$.\cite{broedersz2010cross}  At low prestress (0.2Pa), corresponding to the linear elastic regime previously identified in figure \ref{fig:stresspulse}, both G' and G'' are nearly constant over all frequencies (figure \ref{fig:freqsweeps}, left) indicating minimal flow, which is  consistent with the minimal creep observed at low stress (figure \ref{fig:creepcurves}). Increasing the steady stress to a magnitude corresponding to $\sigma_0$ (figure \ref{fig:freqsweeps}, centre) and beyond to $\sigma^*$ (figure \ref{fig:freqsweeps}, right) reveals a convergence between G''  and G' at the lowest frequencies, indicating the onset of flow. Therefore, crosslink unbinding can be said to occur only over long timescales (exceeding the experimentally accessible time of 1000s) with increases in applied stress leading to faster unbinding kinetics, a phenomenon often referred to as ``slip bond behaviour''.\cite{bell1978models}

We have established that inelastic fluidization occurs above a threshold shear stress of $\sigma^*=2.5\times\sigma_0$ (figure \ref{fig:stressramps}) corresponding to expected filament forces in the piconewton range (figure S4) at unbinding rates of order 10$^{-3}$Hz (figure \ref{fig:freqsweeps}). To verify that inelastic fluidization is indeed caused by crosslink unbinding we seek to artificially modify the affinity and timescale of crosslinking. Firstly, we assemble the filaments in the presence of glutaraldehyde, which induces permanent bonding through fixation. \cite{damink1995glutaraldehyde} To minimize the risk of damaging\cite{lehrer1981damage} or substantially stiffening the filaments through fixation\cite{licup2015stress} we use low glutaraldehyde concentrations of 0.001$\%$(w/v)$\approx 100\mu M$, a 5$\times$ molar excess to vimentin. Fixed networks were found to have comparable loss and storage modulus to untreated networks, validating our approach (figure S5, middle and right panels). We also note previous reports that vimentin intermediate filaments retain their ability to axially stretch even when fixed with glutaraldehyde concentrations as high as 0.1$\%$(w/v)\cite{guzman2006exploring}. By contrast crosslinking is also suppressed by using a  vimentin ``tailless'' mutant, which still leads to self-supporting networks  (figure S5, left panel) a likely consequence of residual crosslinking provided by vimentin's rod domain\cite{pawelzyk2014attractive} and possibly the effect of polymer entanglements.
\begin{figure}[ht]\centering 
\includegraphics[width=1\linewidth]{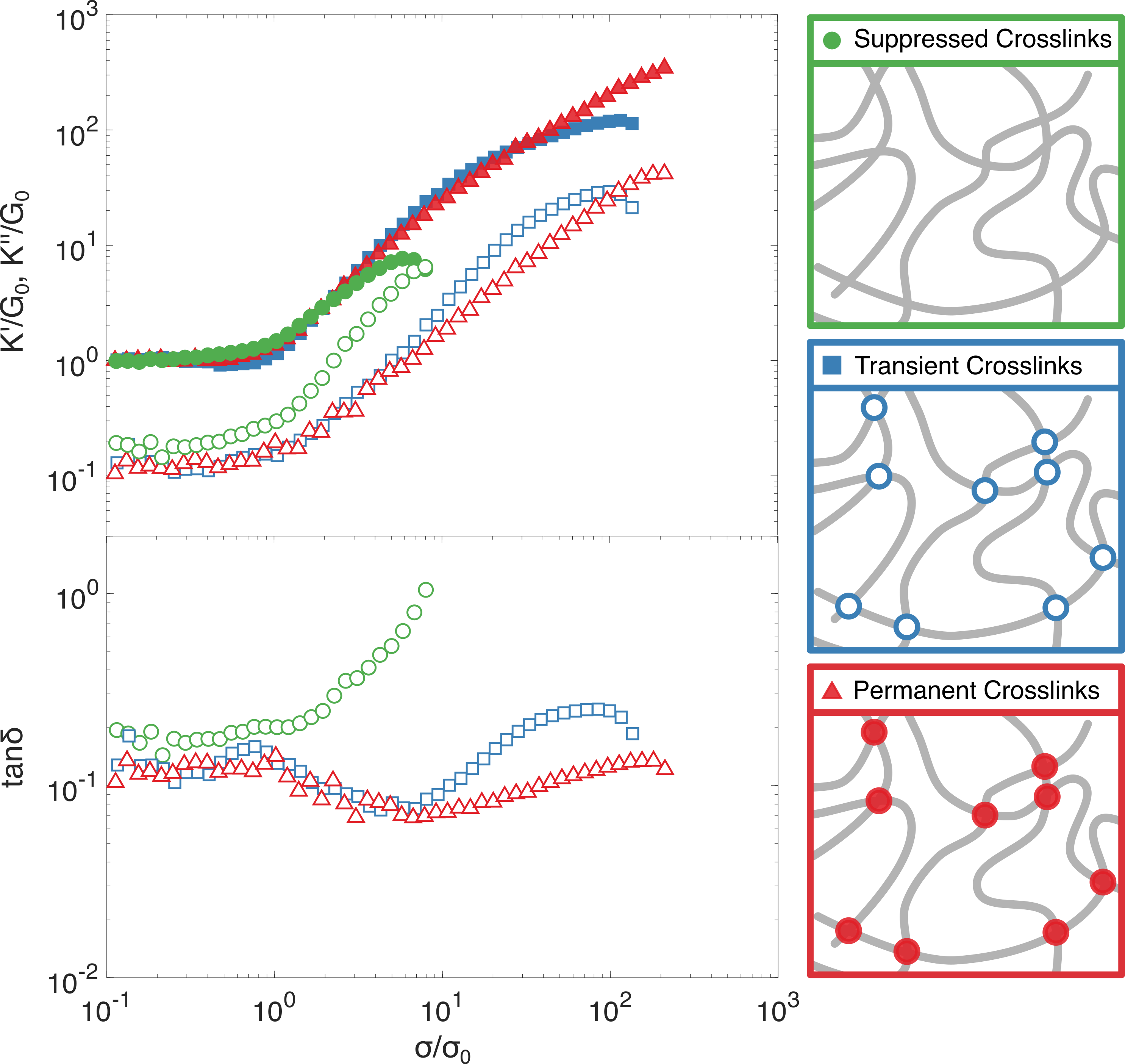}
\caption{Nonlinear mechanics of vimentin networks probed by the stress pulse protocol in different crosslink regimes. Symbols denote transient  crosslinking (squares), permanent crosslinking (triangles) and suppressed crosslinking (circles), closed symbols denote storage modulus, open symbols denote loss modulus.  All networks exhibit stiffening under prestress. Peak modulus and rupture stress increase when  permanent crosslinking is present and decrease when crosslinking is suppressed. The evolution of tan$\delta$ with stress indicates that permanent crosslinking results in lower network flow, while suppressing crosslinking leads to greatly increased network flow. }
\label{fig:differentcrosslinks_pulse}
\end{figure}

\begin{figure}[ht]\centering 
\includegraphics[width=0.85\linewidth]{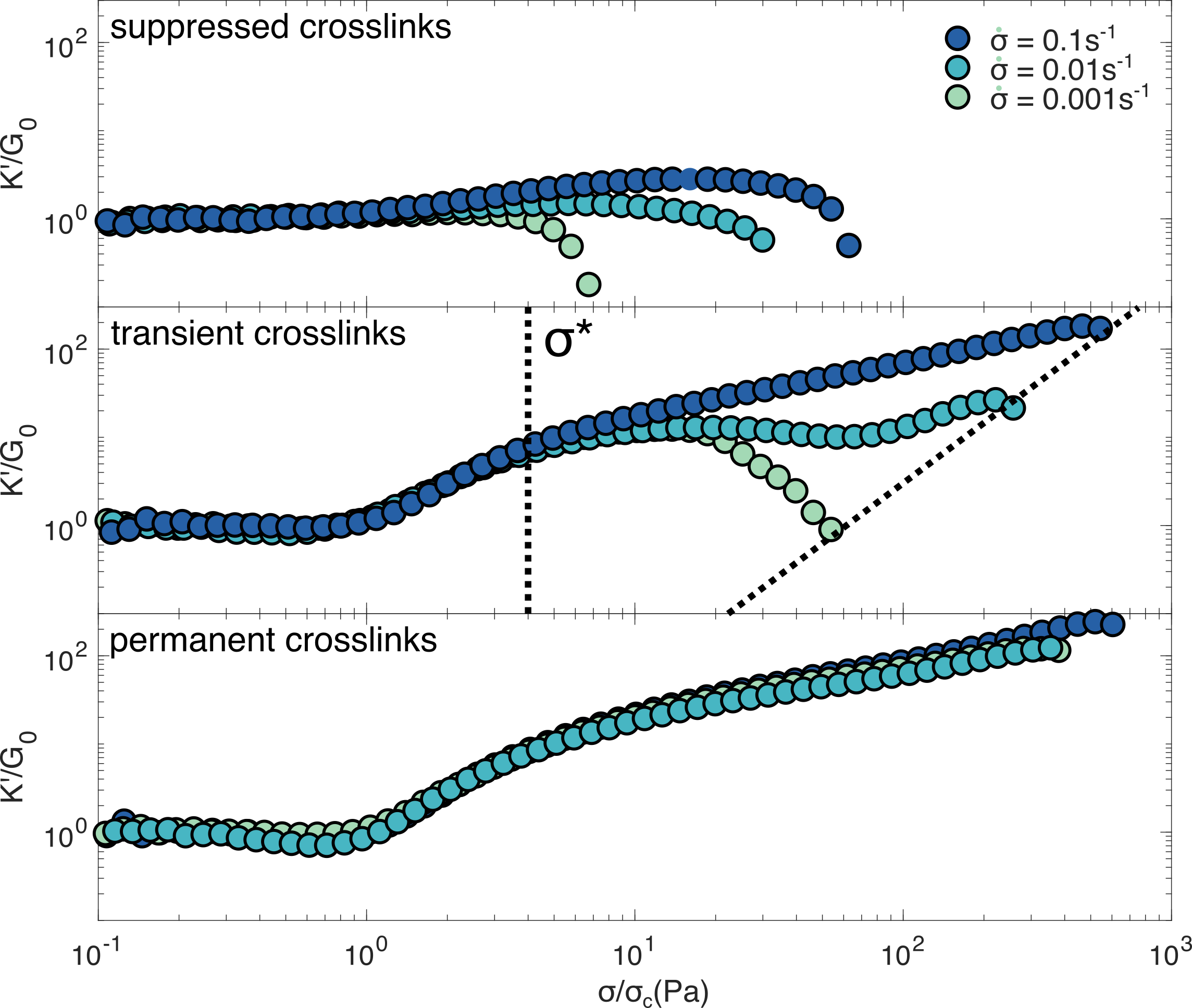}
\caption{Nonlinear mechanics of vimentin networks probed by the stress ramp protocol in different crosslink regimes. Colours denote different loading rates. In the original transiently crosslinked state (centre) loading rate dependent stiffening is observed.  Introducing permanent crosslinks to the network results in a nonlinear response equivalent to transient networks for the fastest loading rates.  Suppressing crosslinking results in virtually no stiffening for all loading rates. }
\label{fig:differentcrosslinks_ramp}
\end{figure}
The nonlinear mechanical response of these three crosslinking regimes (transient, permanent and suppressed crosslinking) using the stress pulse protocol are shown in figure \ref{fig:differentcrosslinks_pulse}. All three are observed to stiffen under applied stress, but the degree of stiffening $K_{max}/G_0$ and the rupture stress $\sigma_{rupture}$ are strongly dependent on the nature of the crosslinking. Permanently crosslinked networks show the most pronounced stiffening and resilience to rupture while  networks with suppressed crosslinking show minimal stiffening and lower rupture stresses. The permanent and transient networks exhibit a similar stiffening response up to stresses of approximately 10$\times\sigma_0$, the same point at which tan$\delta$ is observed to increase in transiently crosslinked networks (figure \ref{fig:differentcrosslinks_pulse}, bottom).  A similar increase in tan$\delta$ is also observed when the crosslinking is suppressed but it occurs at  significantly lower applied stress ($\sigma\approx0.2\times\sigma_0$), indicating a higher propensity for forced crosslink unbinding. Taken together these observations validate our hypothesis that force-dependent crosslink unbinding is the primary factor in determining inelastic fluidization. 

Prior work has shown that tailless vimentin forms thicker and more irregular filaments compared to the full-length protein,\cite{herrmann1996structure,brennich2019mutation} which would result in coarser networks at equivalent concentrations.  To test the influence of this, we increase the concentration of tailless vimentin so that the filament contour length per unit volume $\rho$ matches that of the wild type network, using the previously reported data on the mass density of the filaments.\cite{herrmann1996structure} For equivalent $\rho$ the linear mechanical response is identical between tailless and wild type vimentin networks (figure S5(a)) while the nonlinear mechanical response quantitatively resembles that of  tailless vimentin at lower concentrations (figure S5 (b)). Importantly, also at equal fiber length density, the tailless vimentin stiffens much less than the full length protein, supporting our hypothesis that the tail domain has minimal influence on the linear modulus but is essential for robust strain stiffening. 

Finally, we examine the loading rate dependent mechanics of the three different networks through the stress ramp protocol (figure \ref{fig:differentcrosslinks_ramp}). As expected, networks where crosslinking is suppressed exhibit minimal stress stiffening but pronounced softening at increasing loads, a phenomenon observed across all loading rates.  By contrast, networks where crosslinking is permanent show high degrees of stiffening,  with  little sensitivity to loading rate.  Thus, we can conclusively state that the loading rate dependent mechanics of vimentin networks is a direct consequence of their transient crosslinking.

\section*{Discussion}
\addcontentsline{toc}{section}{Discussion}
The overarching theme to our findings is that vimentin intermediate filament networks are not permanent structures but are dynamic as a consequence of the transient noncovalent interactions between the filaments. Similar characteristics have been observed in the other filamentous systems of the cytoskeleton, actin\cite{ward2008dynamic} and microtubules\cite{lin2007viscoelastic}. In the case of these other cytoskeletal networks, inelastic fluidization appears to counteract their ability to withstand mechanical loads. By contrast, we show that vimentin networks are able to withstand much higher strains before rupture but also possess the ability to soften, counteracting the entropic stiffening. 

Inelastic fluidization in intermediate filament networks is manifested by a characteristic yield stress $\sigma^*$, above which crosslink unbinding is enabled, and the network becomes highly sensitive to loading rate. This yield stress, defined by the onset of rate dependence in the stress pulse protocol (figure \ref{fig:stressramps}) lies comfortably within the range of contractile forces that are typically exerted by cells, of order 0.1-10Pa.\cite{jansen2013cells} Furthermore the networks do not permanently fracture since self-healing is observed upon removal of stress. 

We estimate from assuming an affine network response that $\sigma^*$ corresponds to forces exerted on individual filaments that are in the range of at most a few piconewtons (figure S4). Although rupture forces between tail domains have yet to be directly measured, the interaction potential between the disordered sidechains of neurofilaments, the intermediate filaments found in neuronal cells, is similarly low, of order of a few $k_BT$.\cite{beck2010gel} These piconewton forces exerted on filaments are apparently sufficient to cause unbinding of the tail-mediated crosslinks.  This implies that the tail-mediated crosslinks are significantly weaker than the crosslinks in actin networks provided by accessory crosslinker proteins, which exhibit typical rupture forces of order 10 pN.\cite{ferrer2008measuring} Nevertheless, reconstituted actin networks are able to withstand shear stresses that are about a factor of 10 lower than vimentin networks.\cite{kasza2010actin}  We speculate that the differences between the strength of vimentin and actin networks may arise because binding between actin and its associated accessory proteins is defined locally by a single surface binding site while intermediate filament crosslinking is mediated by the collective interactions between their tail domains. Although these interactions are individually weaker, they may impart a stronger load resistance through a ``strength in numbers'' mechanism. Such a mechanism would suggest slower unbinding rates because unbinding would demand a collective rupture of all tail domains. Indeed we observe crosslink unbinding at particularly slow rates, ($\approx$1000s at $\sigma\approx\sigma^*$), compared to the faster ($\approx$10s) unbinding timescales of actin crosslinkers.\cite{mulla2019frustrated} A further possible factor in vimentin's slow unbinding is that unbound tail domains will maintain a close proximity to each other if neighbouring tail domains remain bound. This would lead to an increased tendency for rebinding, in contrast to actin networks where accessory proteins are able to rapidly diffuse and bind elsewhere in the network. \cite{mulla2018crosslinker,mulla2019frustrated} The slow dynamics we observe for vimentin filaments are consistent with the physiological role of intermediate filaments as maintainers of cell integrity\cite{seltmann2013keratins,harris2012characterizing,latorre2018active} and cytoskeletal polarity in migrating cells\cite{gan2016vimentin,jiu2017vimentin,de2018intermediate} in direct contrast to actin's role in ``fast'' cellular processes such as motility and shape change. \cite{lodish1995molecular}
 
\begin{figure}[ht]\centering 
\includegraphics[width=1\linewidth]{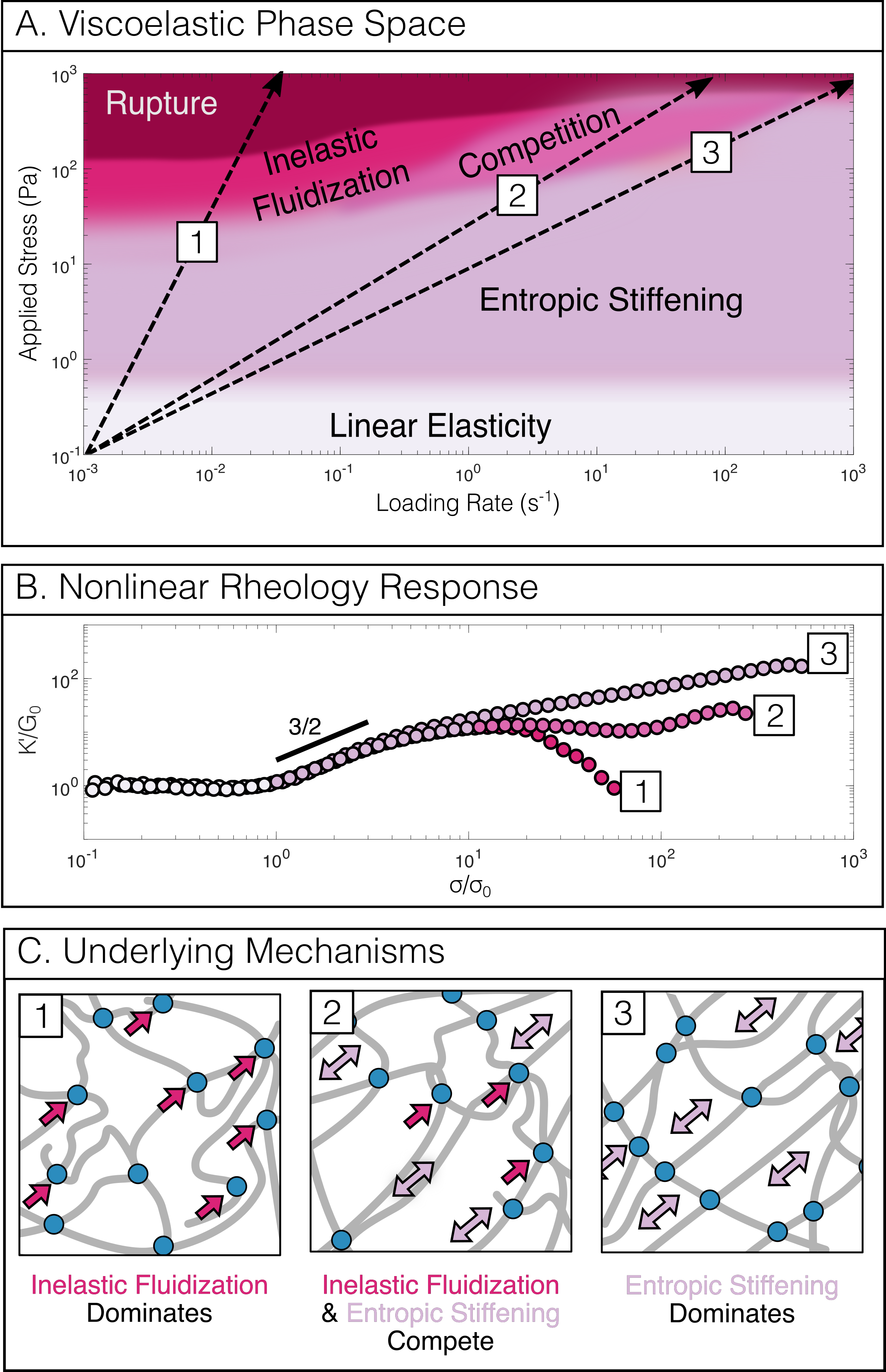}
\caption{Schematic of nonlinear response of transiently crosslinked vimentin networks (a) based on rheological data (b). The interplay between entropic stiffening of filament segments between crosslinks (light arrows) and the dynamic binding/unbinding of crosslinks between them (dark arrows) leads to a nonlinear mechanical response that is sensitive both to the \textit{magnitude} and the \textit{rate} of the applied force (c). For the slowest loading rates (1), the rheological response is dominated by crosslink unbinding, while at the fastest loading rates (3) it is dominated by entropic stiffening. At intermediate rates (2), entropic stiffening and crosslink unbinding occur concurrently such that softening followed by reentrant stiffening is observed. }
\label{fig:statediagram}
\end{figure}

Based on our rheological data we propose a constitutive phase space for the nonlinear mechanics of intermediate filament networks (figure \ref{fig:statediagram}).  This phase space reflects a balance between two phenomena: entropic stiffening at fast loading rates and low applied stress, and inelastic fluidization at slow loading rates and high applied stress. It has previously been proposed, that either one or the other phenomenon will dominate, depending on the relevant experimental time scale.\cite{wolff2012resolving}  The stiffening we observe is nearly instantaneous, occurring over the $\sim$20s timescale of the stress pulse protocol (figure \ref{fig:stresspulse}), in direct contrast to the network's slow inelastic fluidization. This large timescale separation enables the networks to exhibit a complex mechanical response in which entropic stiffening and inelastic fluidization act concurrently such that gradual crosslink unbinding can take place while filament segments between crosslinks continue to stiffen.

\section*{Conclusions}
We have shown that intermediate filaments possess a complex mechanical response that depends both on the magnitude of applied stress and its timescale. By  systematically probing the viscoelastic response as a function of load and loading rate, we highlight a number of previously unreported aspects of intermediate filament network mechanics. We show that the networks have a characteristic yield stress, above which the network softens in a loading-rate dependent manner.  Above this yield stress, softening is highly sensitive to loading rate, a phenomenon that is shown to be a direct consequence of vimentin tail-mediated crosslinking.  This crosslinking is transient but has a particularly slow unbinding rate, which we attribute to the collective binding of vimentin's tail domains. The slow unbinding rate may explain why previous studies have indicated that intermediate filament crosslinking is permanent in nature.\cite{lin2010origins} 

As well as being of fundamental interest, these slow unbinding rates may be of direct relevance to understanding force dissipation in the cytoskeleton.  Previous work has attributed this dissipation to phenomena on the single filament level such as changes in protein secondary structure\cite{fudge2003mechanical}, slippage of subunits\cite{mucke2004assessing} and tail-domain unfolding.\cite{kreplak2005exploring} Our findings highlight that considerations at the network level must also be taken into account. 

In the context of vimentin's role in cell mechanics, it will be of particular interest to examine how crosslinking is regulated by the different concentrations of monovalent and divalent ions, which are known to influence the structure of the crosslink-mediating tail regions\cite{brennich2014impact} and their interaction strength.\cite{lin2010origins} and by post-translational modifications such as phosphorylation of the tail domain.\cite{hyder2008providing} In addition it will be interesting to examine the interplay between noncovalent tail-mediated interactions and mobile crosslinkers by introducing reconstituted plectin into intermediate filament networks.\cite{martin2016both}

\section*{Acknowledgments} 
We thank Chase Broedersz, Fred Mackintosh, Sarah K{\"o}ster, Yuval Mulla and Bart Vos for many helpful discussions. We also thank Harald Herrmann for providing vimentin plasmids and for providing guidance with vimentin purification. We thank Marjolein Kuit-Vinkenoog and Jeffrey den Haan for help with protein purification and characterisation. We gratefully acknowledge funding from the Netherlands Organisation for Scientific Research (NWO), with a program grant from the Foundation for Fundamental Research on 
 (FOM Program
grant nr 143). 



\bibliography{sample} 

\begin{thebibliography}{10}

\bibitem{beck2010gel}
R.~Beck, J.~Deek, J.~B. Jones, and C.~R. Safinya.
\newblock Gel-expanded to gel-condensed transition in neurofilament networks
  revealed by direct force measurements.
\newblock {\em Nature materials}, 9(1):40, 2010.

\bibitem{bell1978models}
G.~I. Bell.
\newblock Models for the specific adhesion of cells to cells.
\newblock {\em Science}, 200(4342):618--627, 1978.

\bibitem{birdsall1983correction}
B.~Birdsall, R.~W. King, M.~R. Wheeler, C.~A. Lewis~Jr, S.~R. Goode, R.~B.
  Dunlap, and G.~C. Roberts.
\newblock Correction for light absorption in fluorescence studies of
  protein-ligand interactions.
\newblock {\em Analytical biochemistry}, 132(2):353--361, 1983.

\bibitem{block2018viscoelastic}
J.~Block, H.~Witt, A.~Candelli, J.~C. Danes, E.~J. Peterman, G.~J. Wuite,
  A.~Janshoff, and S.~K{\"o}ster.
\newblock Viscoelastic properties of vimentin originate from nonequilibrium
  conformational changes.
\newblock {\em Science advances}, 4(6):eaat1161, 2018.

\bibitem{block2017nonlinear}
J.~Block, H.~Witt, A.~Candelli, E.~J. Peterman, G.~J. Wuite, A.~Janshoff, and
  S.~K{\"o}ster.
\newblock Nonlinear loading-rate-dependent force response of individual
  vimentin intermediate filaments to applied strain.
\newblock {\em Physical review letters}, 118(4):048101, 2017.

\bibitem{brennich2019mutation}
M.~Brennich, U.~Vainio, T.~Wedig, S.~Bauch, H.~Herrmann, and S.~K{\"o}ster.
\newblock Mutation-induced alterations of intra-filament subunit organization
  in vimentin filaments revealed by saxs.
\newblock {\em Soft matter}, 2019.

\bibitem{brennich2014impact}
M.~E. Brennich, S.~Bauch, U.~Vainio, T.~Wedig, H.~Herrmann, and S.~K{\"o}ster.
\newblock Impact of ion valency on the assembly of vimentin studied by
  quantitative small angle x-ray scattering.
\newblock {\em Soft Matter}, 10(12):2059--2068, 2014.

\bibitem{broedersz2010cross}
C.~P. Broedersz, M.~Depken, N.~Y. Yao, M.~R. Pollak, D.~A. Weitz, and F.~C.
  MacKintosh.
\newblock Cross-link-governed dynamics of biopolymer networks.
\newblock {\em Physical review letters}, 105(23):238101, 2010.

\bibitem{broedersz2010measurement}
C.~P. Broedersz, K.~E. Kasza, L.~M. Jawerth, S.~M{\"u}nster, D.~A. Weitz, and
  F.~C. MacKintosh.
\newblock Measurement of nonlinear rheology of cross-linked biopolymer gels.
\newblock {\em Soft Matter}, 6(17):4120--4127, 2010.

\bibitem{broedersz2014modeling}
C.~P. Broedersz and F.~C. MacKintosh.
\newblock Modeling semiflexible polymer networks.
\newblock {\em Reviews of Modern Physics}, 86(3):995, 2014.

\bibitem{burla2018biopolymers}
F.~Burla, Y.~Mulla, B.~E. Vos, A.~Aufderhorst-Roberts, and G.~H. Koenderink.
\newblock From mechanical resilience to active material properties in
  biopolymer networks.
\newblock {\em Nature Reviews Physics}, mar 2019.

\bibitem{chen2010fluidization}
C.~Chen, R.~Krishnan, E.~Zhou, A.~Ramachandran, D.~Tambe, K.~Rajendran, R.~M.
  Adam, L.~Deng, and J.~J. Fredberg.
\newblock Fluidization and resolidification of the human bladder smooth muscle
  cell in response to transient stretch.
\newblock {\em PloS one}, 5(8):e12035, 2010.

\bibitem{coulombe2000ins}
P.~A. Coulombe, O.~Bousquet, L.~Ma, S.~Yamada, and D.~Wirtz.
\newblock The ‘ins’ and ‘outs’ of intermediate filament organization.
\newblock {\em Trends in cell biology}, 10(10):420--428, 2000.

\bibitem{damink1995glutaraldehyde}
L.~O. Damink, P.~J. Dijkstra, M.~Van~Luyn, P.~Van~Wachem, P.~Nieuwenhuis, and
  J.~Feijen.
\newblock Glutaraldehyde as a crosslinking agent for collagen-based
  biomaterials.
\newblock {\em Journal of materials science: materials in medicine},
  6(8):460--472, 1995.

\bibitem{de2018intermediate}
C.~De~Pascalis, C.~P{\'e}rez-Gonz{\'a}lez, S.~Seetharaman, B.~Bo{\"e}da,
  B.~Vianay, M.~Burute, C.~Leduc, N.~Borghi, X.~Trepat, and
  S.~Etienne-Manneville.
\newblock Intermediate filaments control collective migration by restricting
  traction forces and sustaining cell--cell contacts.
\newblock {\em J Cell Biol}, 217(9):3031--3044, 2018.

\bibitem{esue2006direct}
O.~Esue, A.~A. Carson, Y.~Tseng, and D.~Wirtz.
\newblock A direct interaction between actin and vimentin filaments mediated by
  the tail domain of vimentin.
\newblock {\em Journal of Biological Chemistry}, 281(41):30393--30399, 2006.

\bibitem{fernandez2006master}
P.~Fern{\'a}ndez, P.~A. Pullarkat, and A.~Ott.
\newblock A master relation defines the nonlinear viscoelasticity of single
  fibroblasts.
\newblock {\em Biophysical journal}, 90(10):3796--3805, 2006.

\bibitem{ferrer2008measuring}
J.~M. Ferrer, H.~Lee, J.~Chen, B.~Pelz, F.~Nakamura, R.~D. Kamm, and M.~J.
  Lang.
\newblock Measuring molecular rupture forces between single actin filaments and
  actin-binding proteins.
\newblock {\em Proceedings of the National Academy of Sciences},
  105(27):9221--9226, 2008.

\bibitem{flitney2009insights}
E.~W. Flitney, E.~R. Kuczmarski, S.~A. Adam, and R.~D. Goldman.
\newblock Insights into the mechanical properties of epithelial cells: the
  effects of shear stress on the assembly and remodeling of keratin
  intermediate filaments.
\newblock {\em The FASEB Journal}, 23(7):2110--2119, 2009.

\bibitem{fudge2003mechanical}
D.~S. Fudge, K.~H. Gardner, V.~T. Forsyth, C.~Riekel, and J.~M. Gosline.
\newblock The mechanical properties of hydrated intermediate filaments:
  insights from hagfish slime threads.
\newblock {\em Biophysical journal}, 85(3):2015--2027, 2003.

\bibitem{gan2016vimentin}
Z.~Gan, L.~Ding, C.~J. Burckhardt, J.~Lowery, A.~Zaritsky, K.~Sitterley,
  A.~Mota, N.~Costigliola, C.~G. Starker, D.~F. Voytas, et~al.
\newblock Vimentin intermediate filaments template microtubule networks to
  enhance persistence in cell polarity and directed migration.
\newblock {\em Cell systems}, 3(3):252--263, 2016.

\bibitem{gardel2006stress}
M.~Gardel, F.~Nakamura, J.~Hartwig, J.~C. Crocker, T.~Stossel, and D.~Weitz.
\newblock Stress-dependent elasticity of composite actin networks as a model
  for cell behavior.
\newblock {\em Physical review letters}, 96(8):088102, 2006.

\bibitem{gardel2004elastic}
M.~Gardel, J.~H. Shin, F.~MacKintosh, L.~Mahadevan, P.~Matsudaira, and
  D.~Weitz.
\newblock Elastic behavior of cross-linked and bundled actin networks.
\newblock {\em Science}, 304(5675):1301--1305, 2004.

\bibitem{gittes1998dynamic}
F.~Gittes and F.~MacKintosh.
\newblock Dynamic shear modulus of a semiflexible polymer network.
\newblock {\em Physical Review E}, 58(2):R1241, 1998.

\bibitem{gralka2015inelastic}
M.~Gralka and K.~Kroy.
\newblock Inelastic mechanics: a unifying principle in biomechanics.
\newblock {\em Biochimica et Biophysica Acta (BBA)-Molecular Cell Research},
  1853(11):3025--3037, 2015.

\bibitem{guo2013role}
M.~Guo, A.~J. Ehrlicher, S.~Mahammad, H.~Fabich, M.~H. Jensen, J.~R. Moore,
  J.~J. Fredberg, R.~D. Goldman, and D.~A. Weitz.
\newblock The role of vimentin intermediate filaments in cortical and
  cytoplasmic mechanics.
\newblock {\em Biophysical journal}, 105(7):1562--1568, 2013.

\bibitem{guzman2006exploring}
C.~Guzman, S.~Jeney, L.~Kreplak, S.~Kasas, A.~Kulik, U.~Aebi, and L.~Forro.
\newblock Exploring the mechanical properties of single vimentin intermediate
  filaments by atomic force microscopy.
\newblock {\em Journal of molecular biology}, 360(3):623--630, 2006.

\bibitem{harris2012characterizing}
A.~R. Harris, L.~Peter, J.~Bellis, B.~Baum, A.~J. Kabla, and G.~T. Charras.
\newblock Characterizing the mechanics of cultured cell monolayers.
\newblock {\em Proceedings of the National Academy of Sciences},
  109(41):16449--16454, 2012.

\bibitem{herrmann2007intermediate}
H.~Herrmann, H.~B{\"a}r, L.~Kreplak, S.~V. Strelkov, and U.~Aebi.
\newblock Intermediate filaments: from cell architecture to nanomechanics.
\newblock {\em Nature reviews Molecular cell biology}, 8(7):562, 2007.

\bibitem{herrmann1999characterization}
H.~Herrmann, M.~H{\"a}ner, M.~Brettel, N.-O. Ku, and U.~Aebi.
\newblock Characterization of distinct early assembly units of different
  intermediate filament proteins.
\newblock {\em Journal of molecular biology}, 286(5):1403--1420, 1999.

\bibitem{herrmann1996structure}
H.~Herrmann, M.~H{\"a}ner, M.~Brettel, S.~A. M{\"u}ller, K.~N. Goldie,
  B.~Fedtke, A.~Lustig, W.~W. Franke, and U.~Aebi.
\newblock Structure and assembly properties of the intermediate filament
  protein vimentin: the role of its head, rod and tail domains.
\newblock {\em Journal of molecular biology}, 264(5):933--953, 1996.

\bibitem{herrmann1992identification}
H.~Herrmann, I.~Hofmann, and W.~W. Franke.
\newblock Identification of a nonapeptide motif in the vimentin head domain
  involved in intermediate filament assembly.
\newblock {\em Journal of molecular biology}, 223(3):637--650, 1992.

\bibitem{herrmann2004isolation}
H.~Herrmann, L.~Kreplak, and U.~Aebi.
\newblock Isolation, characterization, and in vitro assembly of intermediate
  filaments.
\newblock In {\em Methods in cell biology}, volume~78, pages 3--24. Elsevier,
  2004.

\bibitem{huber2015cytoskeletal}
F.~Huber, A.~Boire, M.~P. Lopez, and G.~H. Koenderink.
\newblock Cytoskeletal crosstalk: when three different personalities team up.
\newblock {\em Current opinion in cell biology}, 32:39--47, 2015.

\bibitem{hyder2008providing}
C.~L. Hyder, H.-M. Pallari, V.~Kochin, and J.~E. Eriksson.
\newblock Providing cellular signposts--post-translational modifications of
  intermediate filaments.
\newblock {\em FEBS letters}, 582(14):2140--2148, 2008.

\bibitem{janmey1991viscoelastic}
P.~A. Janmey, U.~Euteneuer, P.~Traub, and M.~Schliwa.
\newblock Viscoelastic properties of vimentin compared with other filamentous
  biopolymer networks.
\newblock {\em The Journal of cell biology}, 113(1):155--160, 1991.

\bibitem{janmey2014polyelectrolyte}
P.~A. Janmey, D.~R. Slochower, Y.-H. Wang, Q.~Wen, and A.~C{\=e}bers.
\newblock Polyelectrolyte properties of filamentous biopolymers and their
  consequences in biological fluids.
\newblock {\em Soft Matter}, 10(10):1439--1449, 2014.

\bibitem{jansen2013cells}
K.~A. Jansen, R.~G. Bacabac, I.~K. Piechocka, and G.~H. Koenderink.
\newblock Cells actively stiffen fibrin networks by generating contractile
  stress.
\newblock {\em Biophysical journal}, 105(10):2240--2251, 2013.

\bibitem{jiu2017vimentin}
Y.~Jiu, J.~Per{\"a}nen, N.~Schaible, F.~Cheng, J.~E. Eriksson, R.~Krishnan, and
  P.~Lappalainen.
\newblock Vimentin intermediate filaments control actin stress fiber assembly
  through gef-h1 and rhoa.
\newblock {\em J Cell Sci}, 130(5):892--902, 2017.

\bibitem{kasza2010actin}
K.~Kasza, C.~Broedersz, G.~Koenderink, Y.~Lin, W.~Messner, E.~Millman,
  F.~Nakamura, T.~Stossel, F.~MacKintosh, and D.~Weitz.
\newblock Actin filament length tunes elasticity of flexibly cross-linked actin
  networks.
\newblock {\em Biophysical journal}, 99(4):1091--1100, 2010.

\bibitem{kollmannsberger2011nonlinear}
P.~Kollmannsberger, C.~T. Mierke, and B.~Fabry.
\newblock Nonlinear viscoelasticity of adherent cells is controlled by
  cytoskeletal tension.
\newblock {\em Soft Matter}, 7(7):3127--3132, 2011.

\bibitem{kreplak2005exploring}
L.~Kreplak, H.~B{\"a}r, J.~Leterrier, H.~Herrmann, and U.~Aebi.
\newblock Exploring the mechanical behavior of single intermediate filaments.
\newblock {\em Journal of molecular biology}, 354(3):569--577, 2005.

\bibitem{krishnan2009reinforcement}
R.~Krishnan, C.~Y. Park, Y.-C. Lin, J.~Mead, R.~T. Jaspers, X.~Trepat,
  G.~Lenormand, D.~Tambe, A.~V. Smolensky, A.~H. Knoll, et~al.
\newblock Reinforcement versus fluidization in cytoskeletal
  mechanoresponsiveness.
\newblock {\em PloS one}, 4(5):e5486, 2009.

\bibitem{latorre2018active}
E.~Latorre, S.~Kale, L.~Casares, M.~G{\'o}mez-Gonz{\'a}lez, M.~Uroz, L.~Valon,
  R.~V. Nair, E.~Garreta, N.~Montserrat, A.~del Campo, et~al.
\newblock Active superelasticity in three-dimensional epithelia of controlled
  shape.
\newblock {\em Nature}, 563(7730):203, 2018.

\bibitem{lehrer1981damage}
S.~S. Lehrer.
\newblock Damage to actin filaments by glutaraldehyde: protection by
  tropomyosin.
\newblock {\em The Journal of cell biology}, 90(2):459--466, 1981.

\bibitem{licup2015stress}
A.~J. Licup, S.~M{\"u}nster, A.~Sharma, M.~Sheinman, L.~M. Jawerth, B.~Fabry,
  D.~A. Weitz, and F.~C. MacKintosh.
\newblock Stress controls the mechanics of collagen networks.
\newblock {\em Proceedings of the National Academy of Sciences},
  112(31):9573--9578, 2015.

\bibitem{lieleg2007cross}
O.~Lieleg and A.~R. Bausch.
\newblock Cross-linker unbinding and self-similarity in bundled cytoskeletal
  networks.
\newblock {\em Physical review letters}, 99(15):158105, 2007.

\bibitem{lieleg2008transient}
O.~Lieleg, M.~M. A.~E. Claessens, Y.~Luan, and A.~Bausch.
\newblock Transient binding and dissipation in cross-linked actin networks.
\newblock {\em Physical review letters}, 101(10):108101, 2008.

\bibitem{lieleg2009cytoskeletal}
O.~Lieleg, K.~Schmoller, M.~M. Claessens, and A.~Bausch.
\newblock Cytoskeletal polymer networks: viscoelastic properties are determined
  by the microscopic interaction potential of cross-links.
\newblock {\em Biophysical journal}, 96(11):4725--4732, 2009.

\bibitem{lin2010divalent}
Y.-C. Lin, C.~P. Broedersz, A.~C. Rowat, T.~Wedig, H.~Herrmann, F.~C.
  MacKintosh, and D.~A. Weitz.
\newblock Divalent cations crosslink vimentin intermediate filament tail
  domains to regulate network mechanics.
\newblock {\em Journal of molecular biology}, 399(4):637--644, 2010.

\bibitem{lin2007viscoelastic}
Y.-C. Lin, G.~H. Koenderink, F.~C. MacKintosh, and D.~A. Weitz.
\newblock Viscoelastic properties of microtubule networks.
\newblock {\em Macromolecules}, 40(21):7714--7720, 2007.

\bibitem{lin2010origins}
Y.-C. Lin, N.~Y. Yao, C.~P. Broedersz, H.~Herrmann, F.~C. MacKintosh, and D.~A.
  Weitz.
\newblock Origins of elasticity in intermediate filament networks.
\newblock {\em Physical review letters}, 104(5):058101, 2010.

\bibitem{lodish1995molecular}
H.~Lodish, A.~Berk, S.~L. Zipursky, P.~Matsudaira, D.~Baltimore, J.~Darnell,
  et~al.
\newblock {\em Molecular cell biology}, volume~3.
\newblock WH Freeman New York, 1995.

\bibitem{lopez2018effect}
C.~G. Lopez, O.~Saldanha, A.~Aufderhorst-Roberts, C.~Martinez-Torres, M.~Kuijs,
  G.~H. Koenderink, S.~K{\"o}ster, and K.~Huber.
\newblock Effect of ionic strength on the structure and elongational kinetics
  of vimentin filaments.
\newblock {\em Soft matter}, 14(42):8445--8454, 2018.

\bibitem{mackintosh1995elasticity}
F.~MacKintosh, J.~K{\"a}s, and P.~Janmey.
\newblock Elasticity of semiflexible biopolymer networks.
\newblock {\em Physical review letters}, 75(24):4425, 1995.

\bibitem{majumdar2018mechanical}
S.~Majumdar, L.~C. Foucard, A.~J. Levine, and M.~L. Gardel.
\newblock Mechanical hysteresis in actin networks.
\newblock {\em Soft matter}, 14(11):2052--2058, 2018.

\bibitem{martin2016both}
I.~Martin, M.~Moch, T.~Neckernuss, S.~Paschke, H.~Herrmann, and O.~Marti.
\newblock Both monovalent cations and plectin are potent modulators of
  mechanical properties of keratin k8/k18 networks.
\newblock {\em Soft matter}, 12(33):6964--6974, 2016.

\bibitem{mucke2004assessing}
N.~M{\"u}cke, L.~Kreplak, R.~Kirmse, T.~Wedig, H.~Herrmann, U.~Aebi, and
  J.~Langowski.
\newblock Assessing the flexibility of intermediate filaments by atomic force
  microscopy.
\newblock {\em Journal of molecular biology}, 335(5):1241--1250, 2004.

\bibitem{mulla2018crosslinker}
Y.~Mulla and G.~H. Koenderink.
\newblock Crosslinker mobility weakens transient polymer networks.
\newblock {\em Physical Review E}, 98(6):062503, 2018.

\bibitem{mulla2019frustrated}
Y.~Mulla, H.~Wierenga, C.~Alkemade, P.~R. ten Wolde, and G.~H. Koenderink.
\newblock Frustrated binding of biopolymer crosslinkers.
\newblock {\em Soft Matter}, 2019.

\bibitem{noding2012intermediate}
B.~N{\"o}ding and S.~K{\"o}ster.
\newblock Intermediate filaments in small configuration spaces.
\newblock {\em Physical review letters}, 108(8):088101, 2012.

\bibitem{omary2004intermediate}
M.~B. Omary, P.~A. Coulombe, and W.~I. McLean.
\newblock Intermediate filament proteins and their associated diseases.
\newblock {\em New England Journal of Medicine}, 351(20):2087--2100, 2004.

\bibitem{pawelzyk2014attractive}
P.~Pawelzyk, N.~M{\"u}cke, H.~Herrmann, and N.~Willenbacher.
\newblock Attractive interactions among intermediate filaments determine
  network mechanics in vitro.
\newblock {\em PLoS One}, 9(4):e93194, 2014.

\bibitem{pourati1998cytoskeletal}
J.~Pourati, A.~Maniotis, D.~Spiegel, J.~L. Schaffer, J.~P. Butler, J.~J.
  Fredberg, D.~E. Ingber, D.~Stamenovic, and N.~Wang.
\newblock Is cytoskeletal tension a major determinant of cell deformability in
  adherent endothelial cells?
\newblock {\em American Journal of Physiology-Cell Physiology},
  274(5):C1283--C1289, 1998.

\bibitem{qin2009hierarchical}
Z.~Qin, L.~Kreplak, and M.~J. Buehler.
\newblock Hierarchical structure controls nanomechanical properties of vimentin
  intermediate filaments.
\newblock {\em PloS one}, 4(10):e7294, 2009.

\bibitem{quinlan2017rim}
R.~A. Quinlan, N.~Schwarz, R.~Windoffer, C.~Richardson, T.~Hawkins, J.~A.
  Broussard, K.~J. Green, and R.~E. Leube.
\newblock A rim-and-spoke hypothesis to explain the biomechanical roles for
  cytoplasmic intermediate filament networks.
\newblock {\em J Cell Sci}, 130(20):3437--3445, 2017.

\bibitem{robert2016intermediate}
A.~Robert, C.~Hookway, and V.~I. Gelfand.
\newblock Intermediate filament dynamics: What we can see now and why it
  matters.
\newblock {\em BioEssays}, 38(3):232--243, 2016.

\bibitem{schmoller2010cyclic}
K.~Schmoller, P.~Fernandez, R.~Arevalo, D.~Blair, and A.~Bausch.
\newblock Cyclic hardening in bundled actin networks.
\newblock {\em Nature communications}, 1:134, 2010.

\bibitem{schopferer2009desmin}
M.~Schopferer, H.~B{\"a}r, B.~Hochstein, S.~Sharma, N.~M{\"u}cke, H.~Herrmann,
  and N.~Willenbacher.
\newblock Desmin and vimentin intermediate filament networks: their
  viscoelastic properties investigated by mechanical rheometry.
\newblock {\em Journal of molecular biology}, 388(1):133--143, 2009.

\bibitem{seltmann2013keratins}
K.~Seltmann, A.~W. Fritsch, J.~A. K{\"a}s, and T.~M. Magin.
\newblock Keratins significantly contribute to cell stiffness and impact
  invasive behavior.
\newblock {\em Proceedings of the National Academy of Sciences}, page
  201310493, 2013.

\bibitem{semmrich2008nonlinear}
C.~Semmrich, R.~J. Larsen, and A.~R. Bausch.
\newblock Nonlinear mechanics of entangled f-actin solutions.
\newblock {\em Soft Matter}, 4(8):1675--1680, 2008.

\bibitem{strelkov2003molecular}
S.~V. Strelkov, H.~Herrmann, and U.~Aebi.
\newblock Molecular architecture of intermediate filaments.
\newblock {\em Bioessays}, 25(3):243--251, 2003.

\bibitem{trepat2007universal}
X.~Trepat, L.~Deng, S.~S. An, D.~Navajas, D.~J. Tschumperlin, W.~T. Gerthoffer,
  J.~P. Butler, and J.~J. Fredberg.
\newblock Universal physical responses to stretch in the living cell.
\newblock {\em Nature}, 447(7144):592, 2007.

\bibitem{wang1994control}
N.~Wang and D.~E. Ingber.
\newblock Control of cytoskeletal mechanics by extracellular matrix, cell
  shape, and mechanical tension.
\newblock {\em Biophysical journal}, 66(6):2181--2189, 1994.

\bibitem{ward2008dynamic}
S.~M.~V. Ward, A.~Weins, M.~R. Pollak, and D.~A. Weitz.
\newblock Dynamic viscoelasticity of actin cross-linked with wild-type and
  disease-causing mutant $\alpha$-actinin-4.
\newblock {\em Biophysical journal}, 95(10):4915--4923, 2008.

\bibitem{wiche2011plectin}
G.~Wiche and L.~Winter.
\newblock Plectin isoforms as organizers of intermediate filament
  cytoarchitecture.
\newblock {\em Bioarchitecture}, 1(1):14--20, 2011.

\bibitem{winheim2011deconstructing}
S.~Winheim, A.~R. Hieb, M.~Silbermann, E.-M. Surmann, T.~Wedig, H.~Herrmann,
  J.~Langowski, and N.~M{\"u}cke.
\newblock Deconstructing the late phase of vimentin assembly by total internal
  reflection fluorescence microscopy (tirfm).
\newblock {\em PLoS One}, 6(4):e19202, 2011.

\bibitem{wolff2012resolving}
L.~Wolff, P.~Fern{\'a}ndez, and K.~Kroy.
\newblock Resolving the stiffening-softening paradox in cell mechanics.
\newblock {\em PloS one}, 7(7):e40063, 2012.

\end{thebibliography}
\bibliographystyle{abbrv} 
\newpage
\onecolumn
\section*{Supplementary Information}
\addcontentsline{toc}{section}{Supplementary Information}

\begin{figure*}[ht]\centering 
\includegraphics[width=0.48\linewidth]{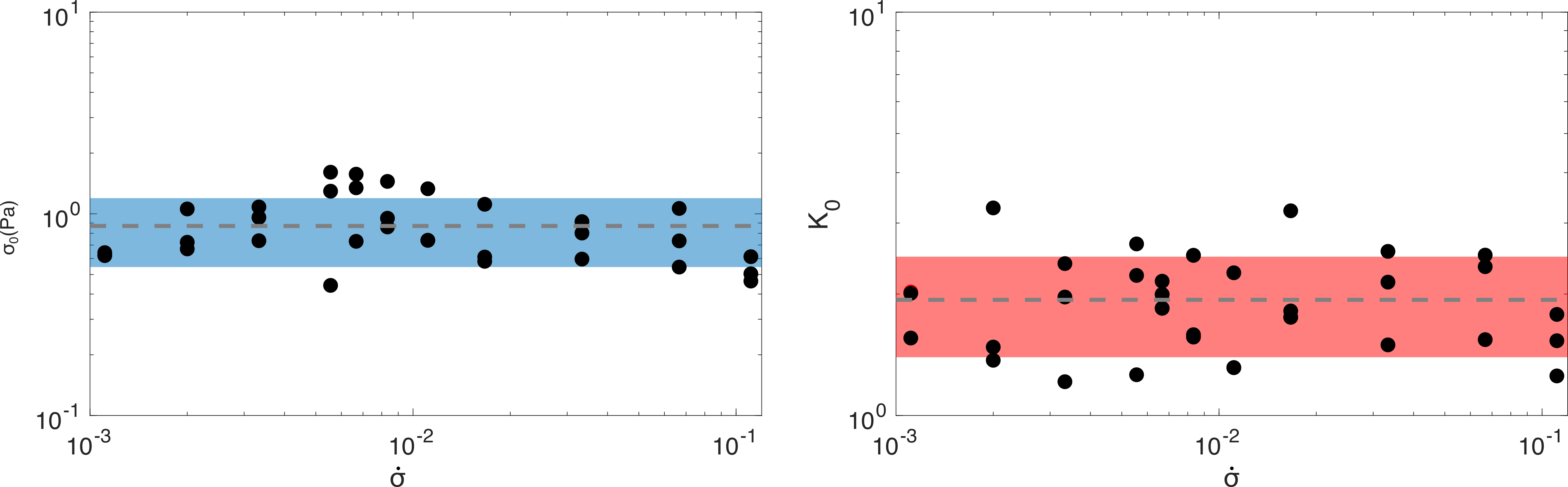}
\captionsetup{labelformat=empty}
    \caption{\textbf{Figure S1}: Onset stress $\sigma_0$  of nonlinearity in the shear moduli of vimentin networks (left) and linear modulus $G_0$ as a function of loading rate. Dashed lines indicate average values, shaded areas indicate an interval of one standard deviation. Unlike the peak modulus and rupture strain (figure \ref{fig:stressramps}) both $G_0$ and $\sigma_0$ are independent of loading rate. }
    \label{fig:k0+sigmacstats}
\end{figure*}

\begin{figure*}[ht]\centering 
\includegraphics[width=0.48\linewidth]{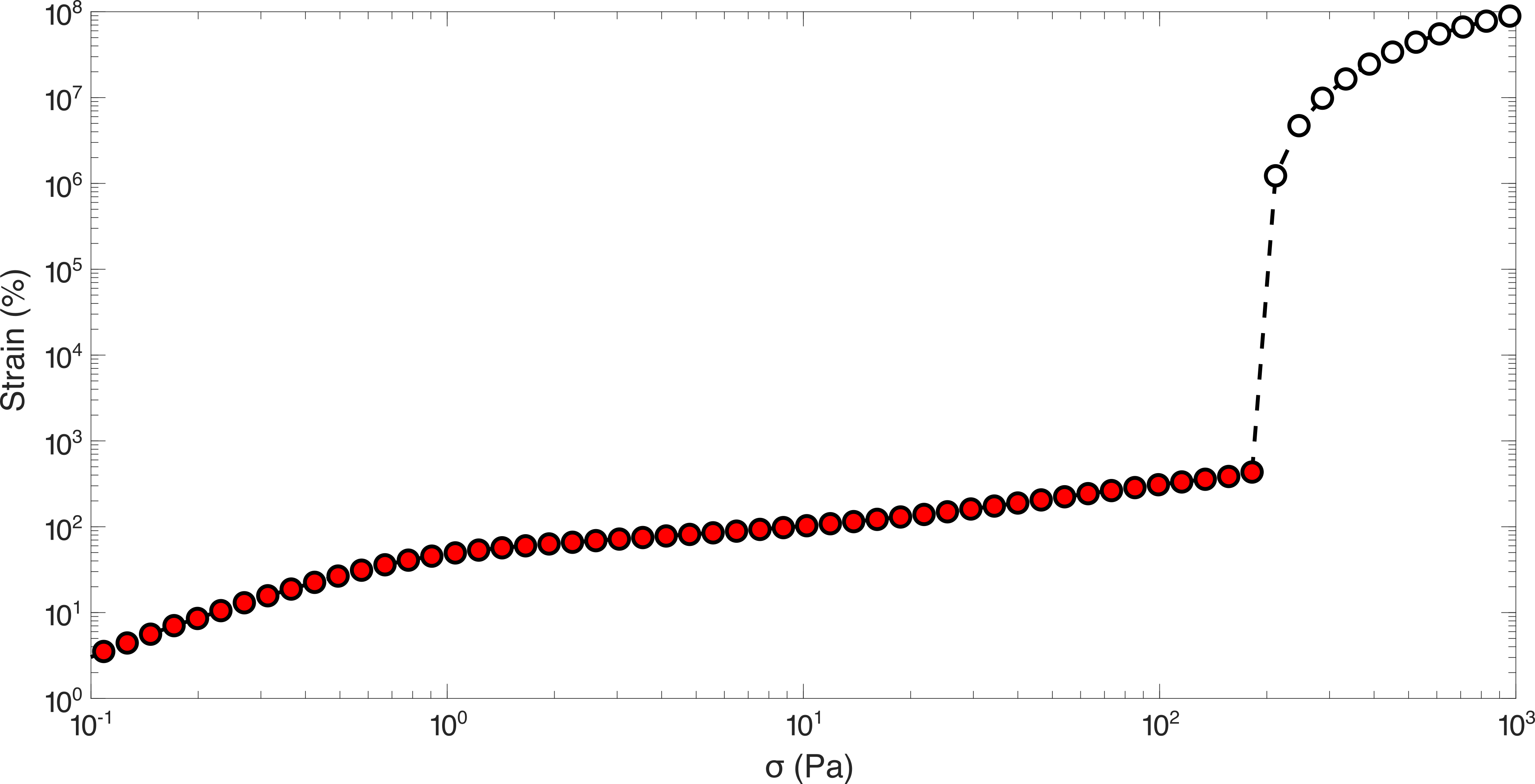}
\captionsetup{labelformat=empty}
    \caption{\textbf{Figure S2:} Typical strain vs. curve of a vimentin network ($\dot{\sigma} = 0.1s^{-1}$) obtained using the stress ramp protocol. Network rupture is defined as having taken place when the strain increases by at least 5 orders of magnitude between successive measurements. Filled symbols show data before rupture, empty symbols show data after rupture. }
    \label{fig:linearfreqsweeps}
\end{figure*}

\begin{figure*}[ht]\centering 
\includegraphics[width=0.48\linewidth]{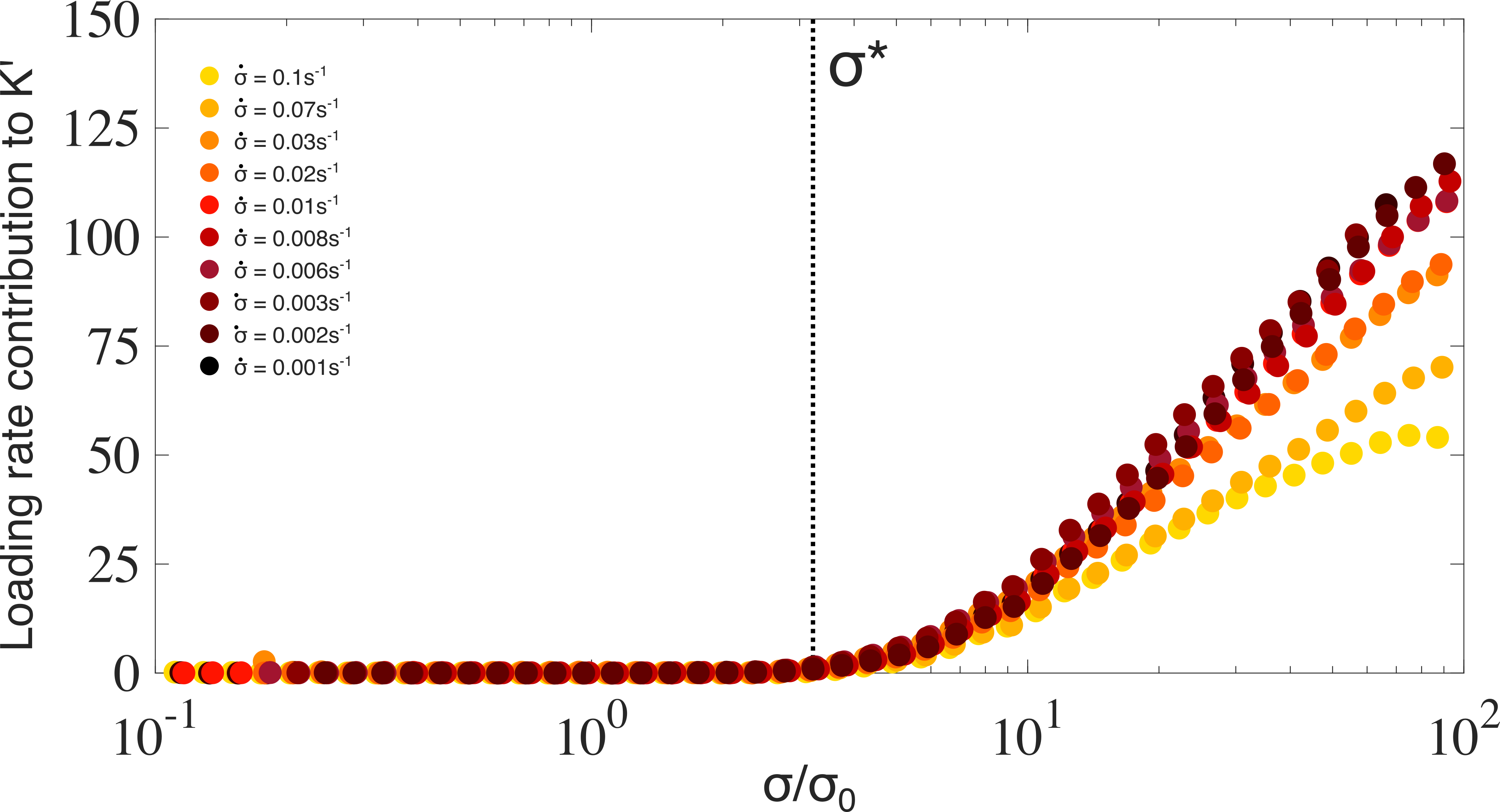}
\captionsetup{labelformat=empty}
    \caption{\textbf{Figure S3:} Loading rate dependent contribution to differential modulus K' as a function of normalized stress. The contribution is calculated by subtracting loading rate dependent nonlinear rheology data (figure \ref{fig:stressramps}) from loading rate \textit{independent} data (figure \ref{fig:stresspulse}). At low applied stress, all data superimposes, and the contribution is negligible. At applied stresses exceeding $\sigma^*$,  loading rate dependent deviations from stress pulse data are observed, with slower loading rates corresponding to larger deviations. }
\end{figure*}

\begin{figure*}[ht]\centering 
\includegraphics[width=0.6\linewidth]{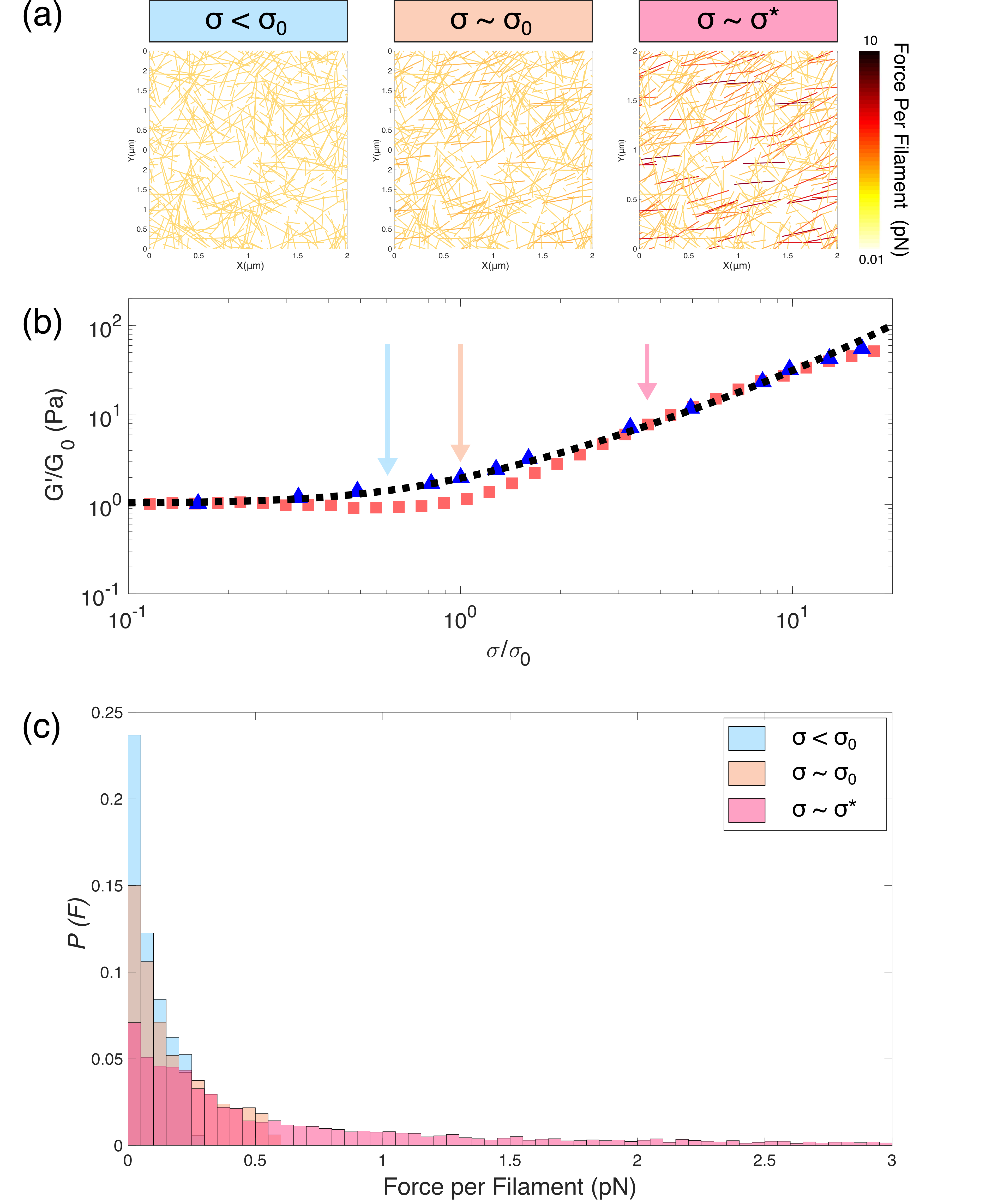}
\captionsetup{labelformat=empty}
    \caption{\textbf{Figure S4:} Simulation of the response of affinely deforming polymer networks under applied shear. (a) Initially isotropic filament ensembles are generated by placing filaments at random positions and orientations and applying a homogeneous shear strain.  The force per filament is calculated using an analytical expression for the force-extension relation of an inextensible semiflexible polymer. (b) The evolution of the resulting storage modulus (black line) is in approximate agreement with the differential storage modulus measured through the stress-pulse protocol of 1mg/ml vimentin (red squares). Discrepancies between simulation and experiment are seen at $\sigma\approx\sigma_0$, which is likely due to slight network nonaffinity as agreement is observed at higher vimentin concentrations of 2mg/ml(blue triangles, taken from\cite{lin2010divalent}).   (c) The force per filament remains below 1pN for all filaments in the linear regime ($\sigma<\sigma_0$) and at the onset of stiffening (($\sigma\approx\sigma_0$). At the yield stress ($\sigma\approx\dot{\sigma}$) the force per filament exceeds 1pN for a small proportion ($\approx15\%$) of filaments.  }
\end{figure*}

\begin{figure*}[ht]\centering 
\includegraphics[width=\linewidth]{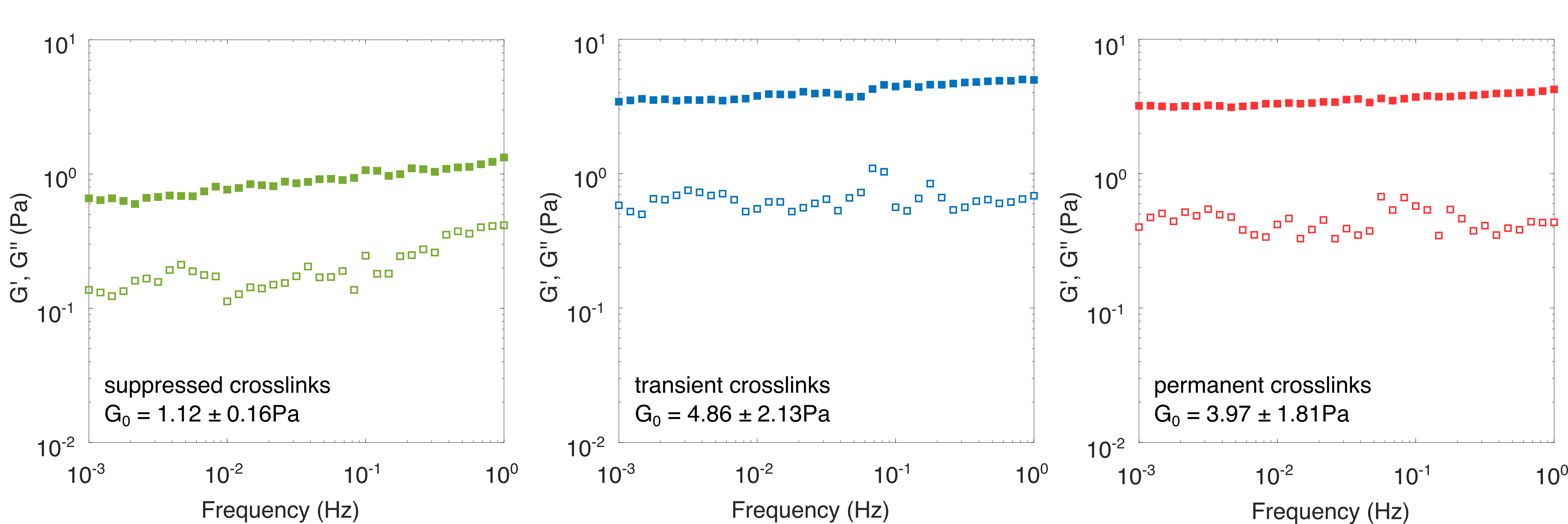}
\captionsetup{labelformat=empty}
    \caption{\textbf{Figure S5:} Frequency sweeps of vimentin networks in different crosslink regimes, averaged (N=3), prepared at concentrations of 1mg/ml, strain amplitude 0.5$\%$. Transiently crosslinked networks have comparable values of  G' and G'' to permanently crosslinked networks, indicating that the glutaraldehyde crosslinking does not significantly alter the network architecture or dynamics.  When crosslinking is suppressed, both G' and G'' are notably lower across all frequencies, which is likely a consequence of their increased mass per length\cite{herrmann1996structure}, resulting in a coarser network at equivalent concentrations. Further evidence for this is presented below in figure S6. }
\end{figure*}

\begin{figure*}[ht]\centering 
\includegraphics[width=\linewidth,trim={0 -12pt 0 0},clip]{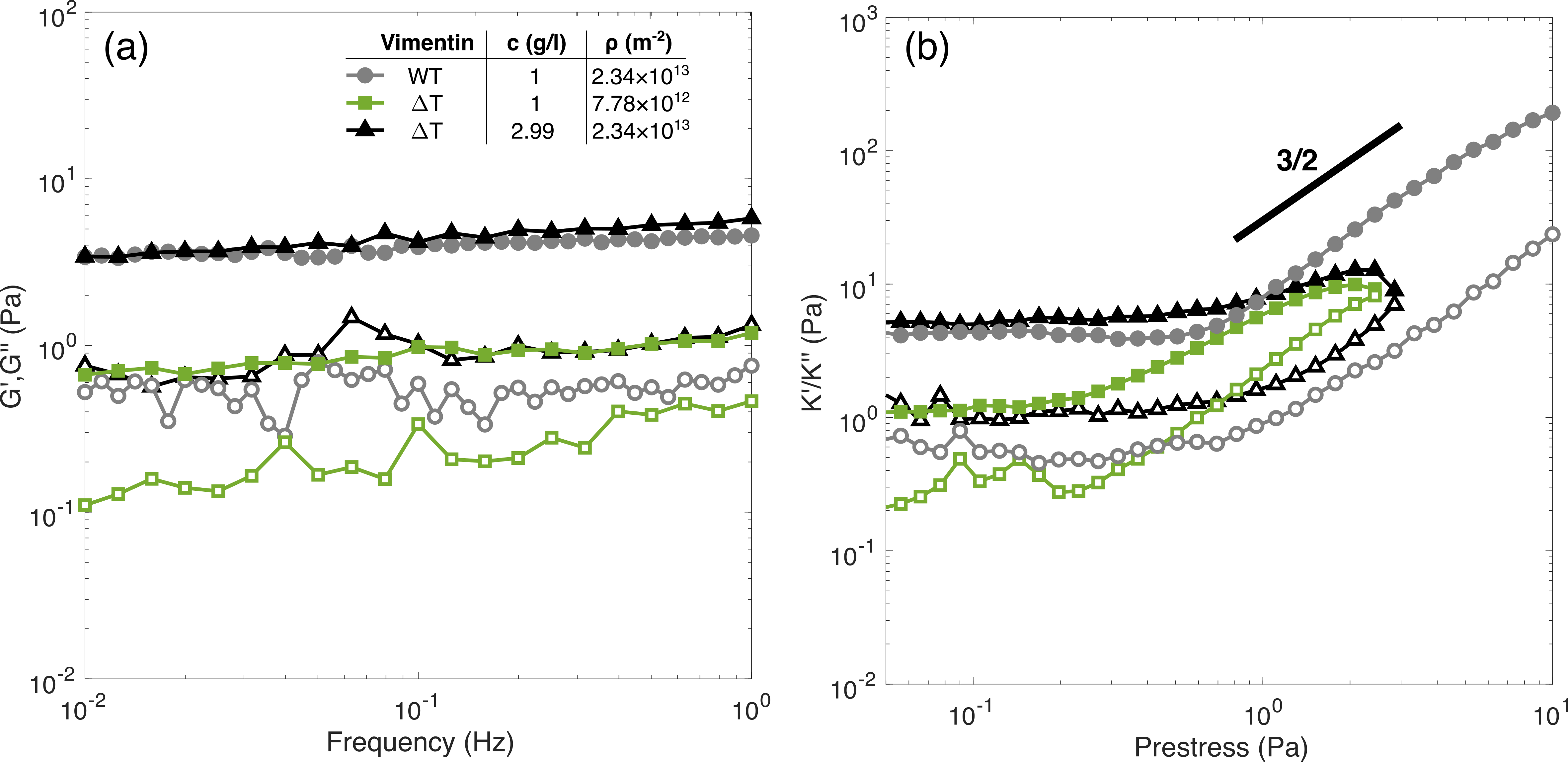}
\captionsetup{labelformat=empty}
    \caption{\textbf{Figure S6:} Linear (a) and nonlinear (b) rheological response of tailless vimentin networks at different concentrations and total filament length per volume, $\rho$. Where the concentration of tailless vimentin is identical to wild type vimentin the linear response is significantly weaker and the degree of stiffening is significantly lower. When both networks have identical $\rho$ their linear response is identical, but the degree of stiffening remains low, clearly demonstrating that crosslinking is the main determinant of reduced stiffening.}
\end{figure*}
\end{document}